\documentclass[prl,aps,twocolumn,longbibliography]{revtex4-1}
\usepackage{color}
\usepackage{graphicx}
\usepackage{bm}
\usepackage{amsmath}
\usepackage{amssymb}
\usepackage{amsthm}
\usepackage{amsfonts}
\usepackage{enumerate}
\usepackage{latexsym}

\usepackage{braket}
\usepackage{array}

\usepackage[colorlinks=true,urlcolor=blue,citecolor=blue,allcolors=blue]{hyperref}
\newcommand{\bs}{\boldsymbol}

\newcommand{\TR}{\text{Tr}}
\newcommand{\TG}{\addtocounter{equation}{1}\tag{\theequation}}

\begin{document}

\title{Entanglement echo and dynamical entanglement transitions}
\author{Kim P\"oyh\"onen}
\author{Teemu Ojanen}
\affiliation{Computational Physics Laboratory, Physics Unit, Faculty of Engineering and
Natural Sciences, Tampere University, P.O. Box 692, FI-33014 Tampere, Finland}
\affiliation{Helsinki Institute of Physics P.O. Box 64, FI-00014, Finland}

\begin{abstract}

We formulate dynamical phase transitions in subsystems embedded in larger quantum systems. Introducing the entanglement echo as an overlap of the initial and instantaneous entanglement ground states, we show its analytic structure after a quench provides natural definition of dynamical phase transitions in the subsystem. These transitions come in two varieties, the entanglement-type transitions and the bulk-type Loschmidt transitions. The entanglement-type transitions arise from periodic reorganization of quantum correlations between the subsystem and its environment, manifesting in instantaneous entanglement ground state degeneracies. Furthermore, the entanglement echo distinguishes the direction of the quench, resolves spatially distinct dynamical phase transitions for non-uniform quenches and give rise to sharply-defined transitions for mixed initial states. We propose an experimental probe to identify entanglement-type transitions through temporal changes in subsystem fluctuations.

\end{abstract}
\maketitle

\emph{Introduction-- }The rapidly growing field of dynamical quantum phase transitions aims to uncover general principles in nonequilibrium many-body dynamics and explore the parallels between dynamics and critical phenomena \cite{Heyl:2013,Karrasch2013,Andraschko2014,Vajna2014,Vajna2015,Budich2016,Zvyagin2016,Halimeh2017}. While there is no direct relation between far-from-equilibrium dynamics and equilibrium phases of matter, recent efforts have revealed a wealth of connections between them \cite{Heyl:2018}. Moreover, in the modern age of quantum simulation and synthetic designer systems, theoretical predictions are directly stimulating new experimental directions \cite{Jurcevic2017,Zhang2017,Flaeschner2018,Guo2019,Tian2019,Wang2019,Tian2020,Xu2020}. Unifying themes across various subfields, dynamical phase transitions have an extraordinarily wide appeal in current research.   

In the present work, we develop the theory of dynamical phase transitions of subsystems of larger many-body systems following a sudden quench, schematically illustrated in Fig.~1 (a). In the diagnostics of quantum correlations in many-body systems, the entanglement spectrum has become an invaluable tool \cite{Li2007,Fidkowski2010}. Recently, it has also found applications in far-from-equilibrium systems \cite{Torlai2014,Canovi2014,Jafari2019,DeNicola2021,gong2018,gong2019,Lu2019,Pastori2020,Surace2020}. Here we consider a bipartite system and introduce the entanglement echo $\mathcal{E}(t)=\langle\lambda_0(0) |\lambda_0(t) \rangle $ as an overlap of the initial and instantaneous ground states $|\lambda_0 \rangle$ of the entanglement Hamiltonian of a subsystem. We show that a vanishing entanglement echo at time $t_c$ provides a natural definition of dynamical phase transitions in a subsystem. The entanglement echo contains essential information on quantum correlations that is not captured by the much-studied Loschmidt echo \cite{Heyl:2013} and signals novel observable properties. 
\begin{figure}[ht]
    \centering
    \includegraphics[width=.99\columnwidth]{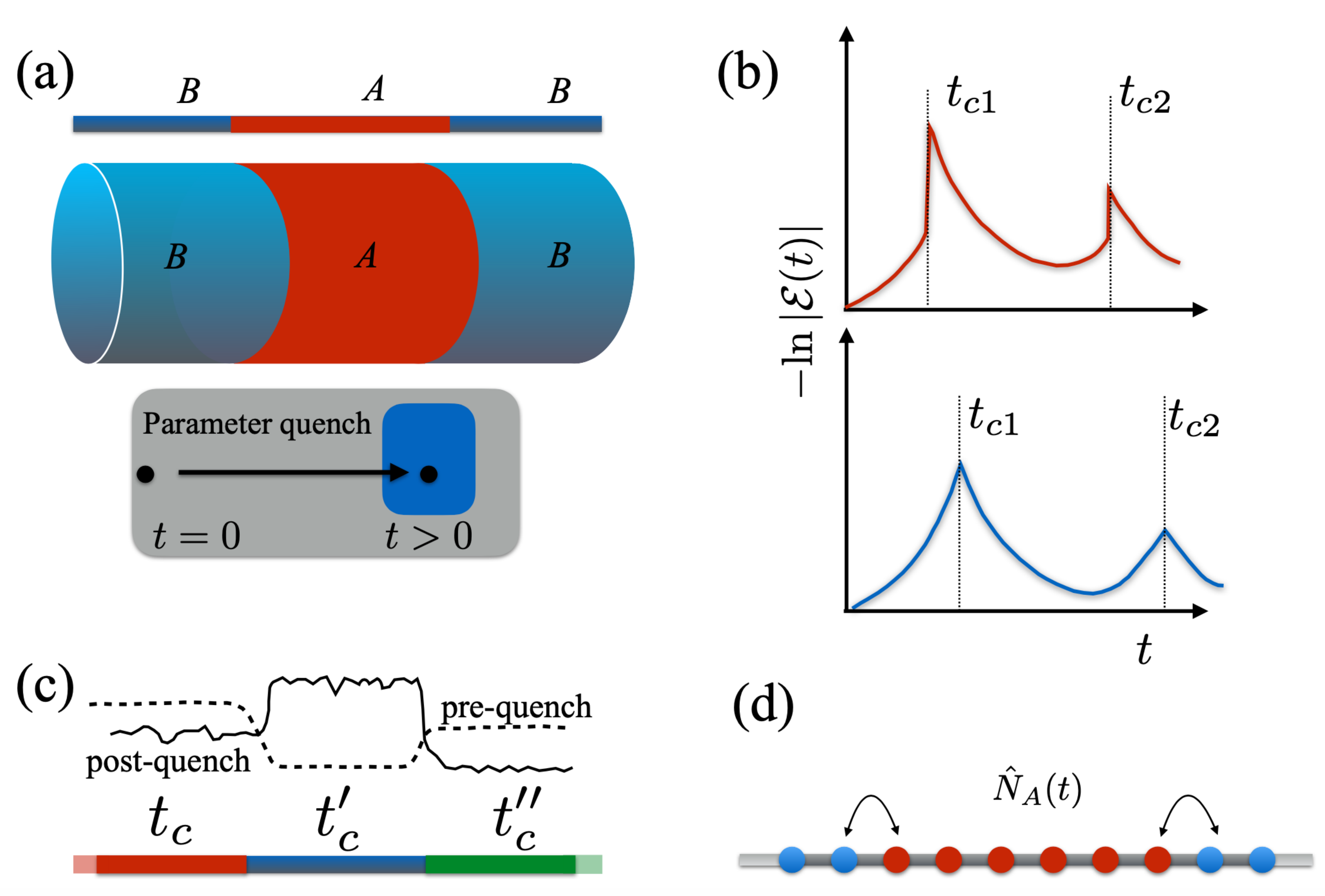}
    \caption{(a): Dynamics of subsystem $A$ embedded in a larger system (seen here for 1d and 2d geometries) display singular features after a sudden quench through a critical point.(b): Entanglement echo $\mathcal{E}(t)$ resolves two types of dynamical criticality, an entanglement-type transition (top) and a bulk-type transition (bottom). (c): Entanglement echo can spatially resolve several dynamical transitions for a single quench configuration (solid and dashed lines). (d): Entanglement-type transitions can be probed by monitoring temporal change in the subsystem observables such as number of particles.  }
    \label{fig1}
\end{figure}

By solving dynamical entanglement transitions in 1d and 2d topological lattice models, we demonstrate substantial conceptual advances in the theory of dynamical phase transitions. Most importantly i) the zeros of the entanglement echo exhibit two types of dynamical criticality as depicted in Fig.~1 (b), the usual Loschmidt-type bulk transitions and the entanglement-type transitions which indicate periodic redistribution of quantum correlations between the subsystems and have no closed system analogy ii) inhomogeneous systems or macroscopically non-uniform quenches give rise to distinct robust spatially-resolved dynamical phase transitions as illustrated in Fig.~1 (c) iii) the entanglement-type transition can be probed by monitoring the temporal behaviour of subsystem fluctuations (Fig.~1 (d)) which also gives rise to oscillating entanglement entropy.  In addition, the entanglement echo provides a natural framework to study entanglement transitions of systems in mixed states and non-unitary evolution.

\emph{Entanglement echo-- }              
To formulate a subsystem's dynamical phase transitions, we consider a bipartioning of a time-evolving system into two subsystems $A$ and $B$. The properties of the subsystem $A$ is encoded in the reduced density matrix $\rho_A(t)=\mathrm{Tr}_B \rho(t)=\sum_i\lambda_i(t)|\lambda_i(t)\rangle\langle\lambda_i(t)|$, obtained by tracing out the degrees of freedom corresponding to the subsystem $B$ from the full density matrix $\rho$ (representing a pure or mixed state). The reduced density matrix can be parametrized by the entanglement Hamiltonian $H_E$ defined by $\rho_A=\frac{e^{-H_E}}{Z}$, where $Z$ ensures the normalization $\mathrm{Tr}_A\,\rho_A=1$. The dominant contribution to $\rho_A$ comes from the state $|\lambda_0\rangle$ with the largest eigenvalue $\lambda_0$, corresponding to the ground state of the entanglement Hamiltonian. The entanglement ground state calculated for a many-body ground state typically encodes universal information about the phase, such as topology and low-lying excitations. The significance of the entanglement ground state points to its potential importance also in far-from-equilibrium systems. Thus, we define the entanglement echo by
\begin{align}\label{eq:echo_general}
\mathcal{E}(t)=\langle\lambda_0(0) |\lambda_0(t) \rangle 
\end{align}
 which measures the overlap between the initial and instantaneous entanglement ground states during temporal evolution. If the entanglement ground state is degenerate in the thermodynamic limit, the echo can be defined as the overlap with the degenerate subspaces. 
 
As depicted in Fig.~1 (a), we consider quench protocols where at $t=0$ the state of the whole system is prepared to a known initial state, such as the ground state or finite-temperature state of a pre-quench Hamiltonian. Then, the Hamiltonian of the system is instantaneously modified to the post-quench form. Analogous to the Loschmidt echo $\mathcal{L}(t)=\langle\Psi(0)|\Psi(t)\rangle$, the vanishing of which defines dynamical phase transitions for the full system, we define dynamical phase transitions for a subsystem in terms of the entanglement echo. We regard the subsystem $A$ as undergoing a dynamical phase transition at time $t_c$ if the entanglement echo vanishes $\mathcal{E}(t_c)=0$. It is convenient to define the entanglement rate function $\Gamma(t)=-\ln |\mathcal{E}(t)|^2/\Omega_A$, where $\Omega_A$ is the characteristic size of the subsystem $A$. Dynamical phase transitions are clearly seen in the non-analytic behaviour of $\Gamma(t)$.

Here we describe two methods of calculating the entanglement echo. In the case of pure initial states and unitary evolution, the state of the system can be expanded $|\Psi(t)\rangle=\sum_{\mu\nu} M_{\mu\nu}(t)|\psi_{\mu}^A\rangle|\psi_{\nu}^B\rangle$, where $|\psi_{\mu}^{A}\rangle, |\psi_{\nu}^B\rangle$ form a complete basis of each subsystem. The singular-value decomposition of matrix $M_{\mu\nu}(t)$ leads to the Schmidt decomposition of a state as 
\begin{align}\label{eq:schmidt}
|\Psi(t)\rangle=\sum_{i}\lambda_i(t)^{1/2}|\lambda_i(t)\rangle|\lambda_i^B(t)\rangle,     
\end{align}
where the sum contains at most min(dim $A$, dim $B$) terms \cite{Ekert1995}. The entanglement ground state at time $t$ can be readily read off  from \eqref{eq:schmidt}, allowing a direct evaluation of the entanglement echo \eqref{eq:echo_general}.

For non-interacting fermions, the evaluation of the entanglement echo simplifies. Pioneered by Peschel \cite{Peschel_2003,Peschel_2009}, the entanglement spectrum for free fermions in a Gaussian state can be obtained from the correlation matrix $\mathcal{C}_{lm}^{\sigma\sigma'}=\langle\hat{c}^\dagger_{l\sigma}\hat{c}_{m\sigma'}\rangle$, where fermion operators $\hat{c}_{m\sigma}$ annihilate particles with spin $\sigma$ and $l,m$ label positions in the subsystem $A$. The eigenstates $|\xi_i \rangle$ and eigenvalues $\xi_i\in [0,1]$ of the correlation matrix can be regarded as the eigenstates  and occupation probabilities of a single-particle entanglement Hamiltonian. To evaluate the entanglement echo, we first need to compute the dynamical correlation matrix  $\mathcal{C}_{lm}^{\sigma\sigma'}(t)=\langle\hat{c}^\dagger_{l\sigma}(t)\hat{c}_{m\sigma'}(t)\rangle$ and diagonalize it. The value $\xi=\frac{1}{2}$ marks the Fermi level of the entanglement Hamiltonian, so the entanglement ground state $|\lambda_0(t)\rangle$ is a Slater determinant constructed from the states satisfying $\frac{1}{2}\leq\xi(t)\leq 1$. In the second-quantized notation, it can be expressed as $|\lambda_0(t)\rangle=\prod_{\xi_i(t)\geq\frac{1}{2}}\hat{c}_{\xi_i(t)}^{\dagger}|0\rangle$. Then, the entanglement echo $\mathcal{E}(t)=\langle\lambda_0(0) |\lambda_0(t) \rangle$ becomes  
 \begin{align}\label{eq:echo_free}
\mathcal{E}(t)=\det{\langle \xi_i(0) |\xi_j(t) \rangle}, 
\end{align}
where the single-particle states satisfy $\xi_i(0),\xi_j(t)\geq \frac{1}{2}$. This formula applies to zero- as well as to finite-temperature pre-quench states.

\emph{Dynamical entanglement transitions in 1d and 2d-- } Now we demonstrate dynamical entanglement phase transitions in solvable two-band Fermi systems. Our analysis applies to arbitrary spatial dimensions but we focus of 1d and 2d topological lattice models. We consider pre- and post quench Hamiltonians of the form  
\begin{equation}\label{eq:twoband}
H^{i/f}=\sum_{l,m}\hat{\bs{c}}_l^\dagger\left[\bs{d}^{i/f}_{lm}\cdot \bs{\sigma}\right]\hat{\bs{c}}_m,    
\end{equation}
where $\bs{\sigma}=(I,\sigma_x,\sigma_y,\sigma_z)$ is a vector of Pauli matrices and the set of matrices $\bs{d}^{i/f}=(d_0^{i/f},d_x^{i/f},d_y^{i/f},d_z^{i/f})$ determine the specific form of pre $(H^{i})$ and post $(H^{f})$ quench Hamiltonians. The spinor operator $\hat{\bs{c}}_l^\dagger=(\hat{c}^{\dagger}_{l\uparrow},\hat{c}^{\dagger}_{l\downarrow})$ creates fermions at site $l$. For translationally invariant systems and quenches, matrices $\bs{d}^{i/f}$ become diagonal in a $n$-dimensional quasimomentum space. With a minor modification, which extends spinors $\hat{\bs{c}}$ to the Nambu space, model \eqref{eq:twoband} also describes quenches in topological superconductors and solvable spin chains \cite{Heyl:2013,Najafi2019,zamani2020} and spin liquids \cite{Schmitt2015}. In Sec.~I of the supplementary material (SI), we have derived the expression for the dynamic correlation matrix for the model \eqref{eq:twoband} for spatially dependent parameters and quench protocols in zero and finite-temperature initial states.     
\begin{figure*}[ht]
    \centering
    \includegraphics[width=1.99\columnwidth]{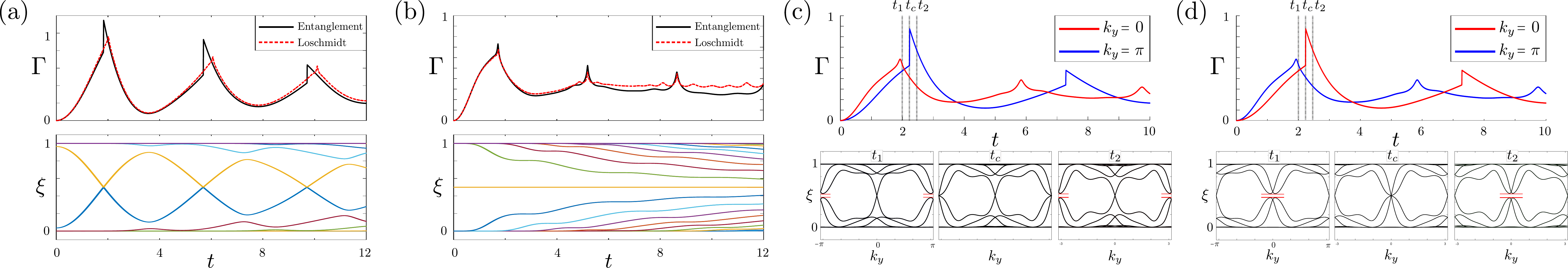}
    \caption{(a): Entanglement echo rate of 1d system (top) for quench $m=1.5 \to m=0.3$ and the system size $L=100$ (total) $L_A=30$ (subsystem). The jump singularities, which arise from the crossing of the entanglement spectrum (bottom), do not coincide with the cusps singularities of the Loschmidt echo of the full system. (b): Same as (a) but for the opposite quench  $m=0.3 \to m=1.5$. The two echos agree (apart from finite size effects which vanish in the thermodynamic limit). (c): Momentum-resolved entanglement echo rate (top) for Chern insulator quench $m=0.5 \to m=-0.5$ and the system size $L=100$ (total) $L_A=30$ (subsystem). The jump singularities correspond to $k_y$ values for which the entanglement spectrum (bottom) exhibits temporal gap closings at $t_c$. (d): Same as (c) but for the opposite quench  $m=-0.5 \to m=0.5$. }
    \label{fig2}
\end{figure*}

We first consider a 1d topological insulator defined by $\bs{d}(k)=(0,\sin k,0,m-\cos k)$. This model belongs to the Altland-Zirnbauer class BDI \cite{Altland1997,Schnyder2008} and exhibits a nontrivial phase for $|m|<1$ and trivial phase for $|m|>1$. In Fig.~2 (a)-(b) we have illustrated the dynamical phase transitions when the system is quenched through the critical point $m=1$. The entanglement echo distinguishes whether the quench is performed from the trivial to the topological phase or vice versa. In the former case, which we dub as an entanglement-type transition, the entanglement echo displays periodic jump discontinuities as seen in shown in Fig.~2 (a). The entanglement spectrum reveals that the jumps arise from stroboscopic level crossing signalling an instantaneous entanglement ground state degeneracy. In the latter case (Fig.~2 (b)), which we call a bulk-type transition, the entanglement echo exhibits cusps and agrees with the Loschmidt rate function. As shown in Sec.~V in the SI, the stroboscopic entanglement ground state degeneracies persists also to finite-temperature initial states. In contrast to the Loschmidt echo, which does not offer a natural generalization with sharply-defined transitions at  finite temperatures \cite{bhattacharya2017, heyl2017,Sedlmayr2018open,abelin2016}, the entanglement-type transitions remain well-defined. While the entanglement echo for the bulk-type transitions reduce to the Loschmidt echo of the total system, as seen in Fig.~2 (a) and discussed in Sec.~III of the SI, the analytic structure and the critical times of entanglement-type transitions do not coincide with the Loschmidt transitions. The entanglement echo quantifies a temporal reorganization of quantum correlations between the two subsystems and captures essential information not contained in the Loschmidt echo.  

Two-dimensional systems exhibit similar bulk- and entanglement-type transitions as 1d systems. By considering the geometry shown in Fig.~1 (a), the subsystem $A$ can be chosen as a segment in the $x$ direction so that the perpendicular momentum $k_y$ remains a good quantum number. The reduced density matrix decouples to blocks labelled by $k_y$, and the entanglement echo can be decomposed from the echoes of each block as $\mathcal{E}(t)=\prod_{k_y} \mathcal{E}(k_y,t)$. The $k_y$-resolved partial echoes can be obtained by diagonalizing the momentum-resolved correlation matrix derived in Sec.~I of the SI. After obtaining the eigenfunctions $|\xi_i(k_y,t)\rangle$, the partial echos can be calculated by applying Eq.~\eqref{eq:echo_free}. In fact, quench dynamics are conveniently analysed in terms of partial echos $\mathcal{E}(k_y,t)$. Here we consider Chern insulators defined by $\bs{d}(k)=(0,\sin k_x,\sin k_y,m-\cos k_x-\cos k_y)$ which exhibits three distinct topological phases with Chern numbers $C=-1$ (when $-2<m<0$), $C=1$ ($0<m<2$) and $C=0$ (when $|m|>2$). As in 1d case, when the system is quenched through a critical point, the entanglement echo shows non-analytic behaviour which depends on the direction of the quench. In addition, the entanglement echo rate function may exhibit either a cusp or jump singularity depending on $k_y$. This is illustrated in Fig.~2 (c)-(d) for transitions between $C=\pm 1$ phases. At times when the rate function shows a jump singularity for specific $k_y$ values, the instantaneous entanglement spectrum exhibits temporal gap closing for the corresponding $k_y$. Also, the momentum for which the gap closing takes place changes when the direction of the quench is inverted. Thus, the entanglement echo in both 1d and 2d systems reveal two distinct dynamical phase transitions and, in contrast to the Loschmidt echo, makes a qualitative distinction in which direction the critical point is crossed.

\emph{Spatially-varying quenches-- }
The Loschmidt echo characterizes dynamics of the system as a whole and, as such, is incapable of providing spatially-resolved information. However, the entanglement echo reveals novel dynamical criticality in macroscopically inhomogeneous systems or spatially varying quenches. In fact, a single quench can give rise to several spatially-resolved dynamical phase transitions characterized by different time scales. In Fig.~3 we have illustrated a quench in 1d system, where the pre- and post-quench configurations vary in space. The subsystem $A$ experiences a quench from a trivial to topological phase while subsystem $B$ experiences the opposite quench. The entanglement echo reveals that, indeed, the different parts of the system exhibit distinct sharply-defined dynamical phase transitions. Not only are their critical times different, but the non-analytic structure shows that the transition in $A$ is of entanglement type and the transition in $B$ is of bulk-type. Since it takes finite time for information to propagate through the system \cite{gong2019}, the short-time behaviour giving rise to early dynamical phase transitions is sensitive only to local quench properties. Thus, a system which exhibits several distinct equilibrium critical points can display multiple spatially-resolved dynamical phase transitions in a single quench.
\begin{figure}[ht]
    \centering
    \includegraphics[width=.6\columnwidth]{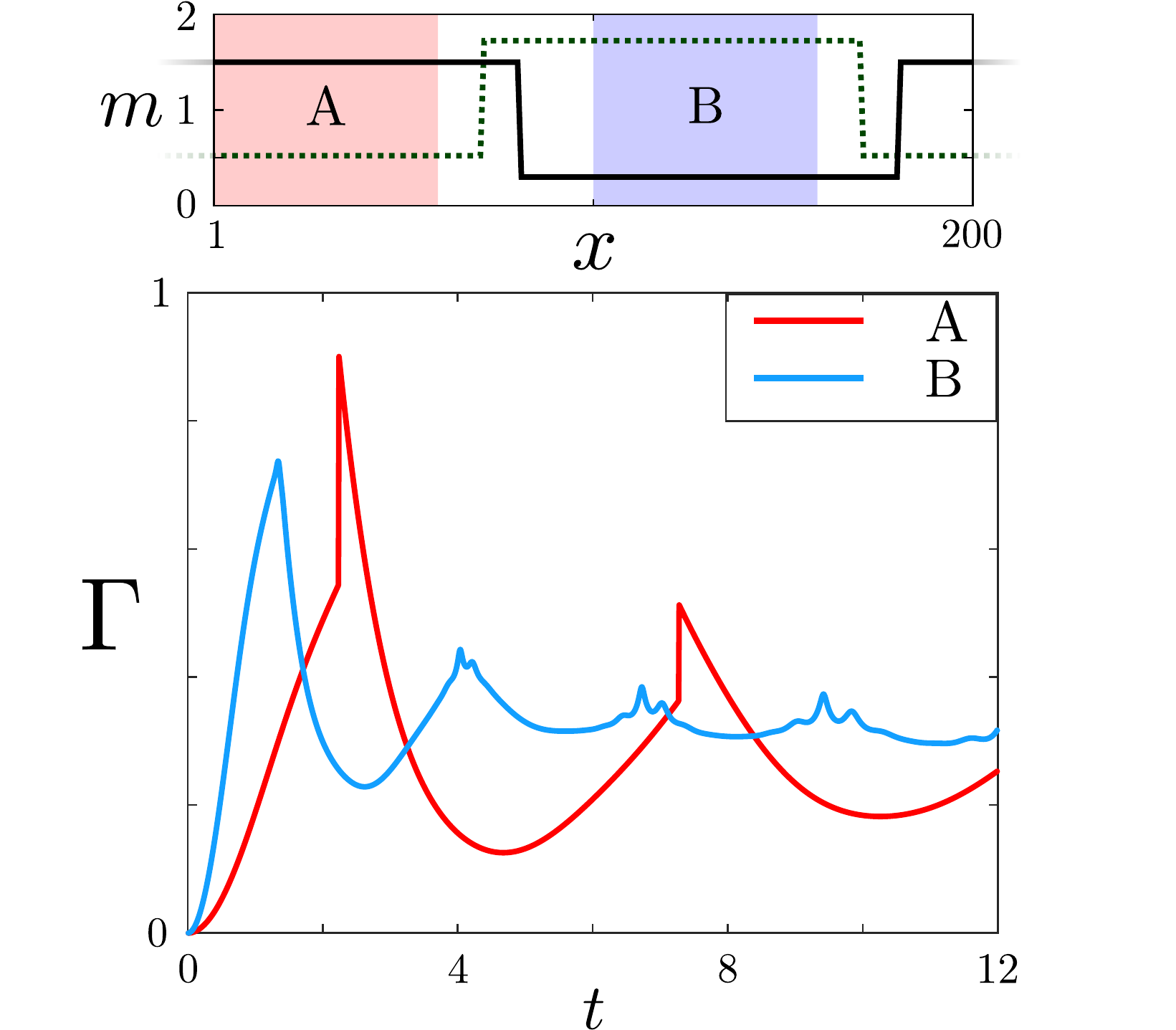}
    \caption{ Top: Spatially-varying pre-quench (solid) and post-quench (dashed) configurations for the 1d model with periodic boundary conditions. The initial configuration consists of regions with $\mu_1 = 1.5$ and $\mu_2 = 0.3$, while the post-quench parameters have $\mu_3 = 0.5$ and $\mu_4 = 1.7$. Bottom: Resulting distinct transitions displayed in subsystems $A$ (red) and $B$ (blue).  }
    \label{fig4}
\end{figure}

\emph{Observable consequences-- }The entanglement-type transitions arise from instantaneous degeneracies of the entanglement ground state which persist to finite-temperature initial states. It is natural to wonder what the observable consequences of this are, especially in contrast to the Loschmidt-type criticality. Far from equilibrium, the states in the entanglement spectrum are not in simple correspondence with the physical edge modes, thus preventing the most direct experimental probes. Here we devise a method to probe and distinguish the entanglement-type transitions by monitoring the temporal changes in subsystem fluctuations. The stroboscopic degeneracy of two entanglement ground states is expected to lead to enhanced fluctuations for observables which have different expectation values in the two states. Indeed, we demonstrate this by considering the number of particles in the subsystem $A$ in a setup depicted in Fig.~1 (d). The particle number operator is $\hat{N}_A=\sum_{i\in A,\alpha=\uparrow,\downarrow}  \hat{c}_{i\alpha}^\dagger\hat{c}_{i\alpha}$, where the summation is over the lattice sites in $A$ and spin. As shown in Sec.~IV of the SI, the time-dependent variance of the particle number is given by
\begin{align} \label{eq:variance2}
&\mathrm{Var}\, N_A(t)=\sum_{i}\left[\xi_i(t)-\xi_i^2(t)\right], 
\end{align}
where $\xi_i(t)$ are the eigenvalues of the correlation matrix. In a translation-invariant system, the first term is a constant fixed by the average density, however, the second term should reflect the pronounced oscillations of midgap states characterizing the entanglement-type transitions shown in Fig.~2 (a). As seen in Fig.~4 (a), the particle number variance indeed oscillates with periodicity of the critical times. In addition to oscillations, it shows a linear trend due to mixing of the two subsystems. The onset time of the linear growth depends on the depth of a quench while the oscillation period reflects the periodicity of critical times. The pronounced oscillations, which are visible even for small subsystems down to $\sim 10$ sites, persist to finite temperature initial states and provide an experimental signal that distinguish entanglement-type transitions from Loschmidt transitions seen in Fig.~4 (b) and trivial quenches shown in Sec.~V in the SI. 
\begin{figure}[ht]
    \centering
    \includegraphics[width=.99\columnwidth]{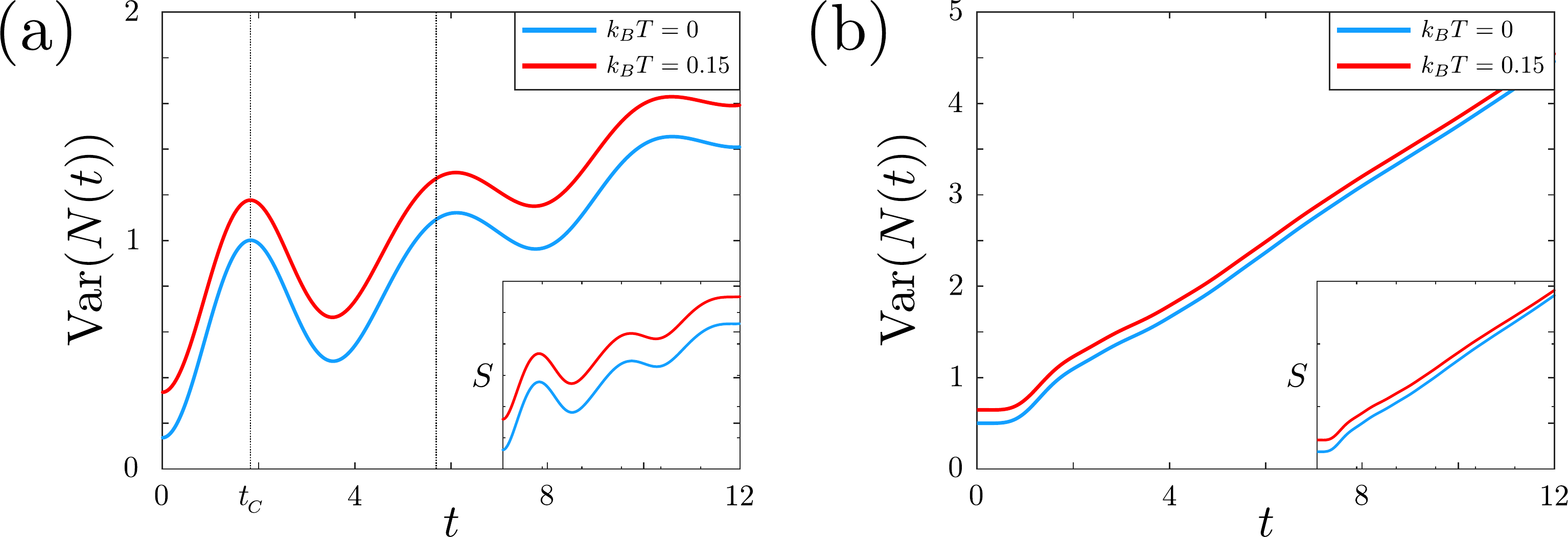}
    \caption{Subsystem particle number variance and entanglement entropy (inset) after a quench. (a) corresponds to the entanglement-type transition in Fig.~1 (a) and (b) to the bulk-type transition in  Fig.~1 (b).  }
    \label{fig4}
\end{figure}
Moreover, as illustrated in Fig.~4, particle number fluctuations essentially reflect the behaviour of the von Neumann entropy $S(t)=-\sum_i\left[\xi_i\log \xi_i+(1-\xi_i)\log (1-\xi_i)\right]$ \cite{Zhang_2014}. The difference in entropy oscillations \cite{Sedlmayr2018} depending on the direction of the quench is naturally explained by the existence of the two types of subsystem transitions discussed in our work. Since the above discussed mechanism of the subsystem fluctuations follow from the oscillating entanglement ground state degeneracy, it applies to generic observables and entanglement-type dynamical phase transitions.

\emph{Conclusion and outlook-- }
In this work we formulated dynamical phase transitions for a subsystem of a many-body system by introducing the entanglement echo. The entanglement echo provides an appropriate generalization of the Loschmidt echo, giving rise to several conceptual advances as well as new observable predictions discussed in our work. In the present work, we studied dynamical criticality resulting from a zero- and finite-temperature initial state undergoing unitary time evolution after a quench. Since the entanglement echo is formulated in terms of the reduced density matrix, it can be straightforwardly employed to study mixed states and non-unitary time evolution.  In the future, it will be interesting to study subsystem dynamics in systems subjected to measurements, the effects of measurements on quench dynamics \cite{kuo2021} and possible measurement-induced dynamical entanglement phase transitions \cite{li2018,Chan2019,skinner2019}.    

\emph{Acknowledgements-- } The authors acknowledge the Academy of Finland project 331094 for support.

\bibliography{biblio}

\begin{thebibliography}{46}%
\makeatletter
\providecommand \@ifxundefined [1]{%
 \@ifx{#1\undefined}
}%
\providecommand \@ifnum [1]{%
 \ifnum #1\expandafter \@firstoftwo
 \else \expandafter \@secondoftwo
 \fi
}%
\providecommand \@ifx [1]{%
 \ifx #1\expandafter \@firstoftwo
 \else \expandafter \@secondoftwo
 \fi
}%
\providecommand \natexlab [1]{#1}%
\providecommand \enquote  [1]{``#1''}%
\providecommand \bibnamefont  [1]{#1}%
\providecommand \bibfnamefont [1]{#1}%
\providecommand \citenamefont [1]{#1}%
\providecommand \href@noop [0]{\@secondoftwo}%
\providecommand \href [0]{\begingroup \@sanitize@url \@href}%
\providecommand \@href[1]{\@@startlink{#1}\@@href}%
\providecommand \@@href[1]{\endgroup#1\@@endlink}%
\providecommand \@sanitize@url [0]{\catcode `\\12\catcode `\$12\catcode
  `\&12\catcode `\#12\catcode `\^12\catcode `\_12\catcode `\%12\relax}%
\providecommand \@@startlink[1]{}%
\providecommand \@@endlink[0]{}%
\providecommand \url  [0]{\begingroup\@sanitize@url \@url }%
\providecommand \@url [1]{\endgroup\@href {#1}{\urlprefix }}%
\providecommand \urlprefix  [0]{URL }%
\providecommand \Eprint [0]{\href }%
\providecommand \doibase [0]{http://dx.doi.org/}%
\providecommand \selectlanguage [0]{\@gobble}%
\providecommand \bibinfo  [0]{\@secondoftwo}%
\providecommand \bibfield  [0]{\@secondoftwo}%
\providecommand \translation [1]{[#1]}%
\providecommand \BibitemOpen [0]{}%
\providecommand \bibitemStop [0]{}%
\providecommand \bibitemNoStop [0]{.\EOS\space}%
\providecommand \EOS [0]{\spacefactor3000\relax}%
\providecommand \BibitemShut  [1]{\csname bibitem#1\endcsname}%
\let\auto@bib@innerbib\@empty
\bibitem [{\citenamefont {Heyl}\ \emph {et~al.}(2013)\citenamefont {Heyl},
  \citenamefont {Polkovnikov},\ and\ \citenamefont {Kehrein}}]{Heyl:2013}%
  \BibitemOpen
  \bibfield  {author} {\bibinfo {author} {\bibfnamefont {M.}~\bibnamefont
  {Heyl}}, \bibinfo {author} {\bibfnamefont {A.}~\bibnamefont {Polkovnikov}}, \
  and\ \bibinfo {author} {\bibfnamefont {S.}~\bibnamefont {Kehrein}},\ }\href
  {\doibase 10.1103/PhysRevLett.110.135704} {\bibfield  {journal} {\bibinfo
  {journal} {Phys. Rev. Lett.}\ }\textbf {\bibinfo {volume} {110}},\ \bibinfo
  {pages} {135704} (\bibinfo {year} {2013})}\BibitemShut {NoStop}%
\bibitem [{\citenamefont {Karrasch}\ and\ \citenamefont
  {Schuricht}(2013)}]{Karrasch2013}%
  \BibitemOpen
  \bibfield  {author} {\bibinfo {author} {\bibfnamefont {C.}~\bibnamefont
  {Karrasch}}\ and\ \bibinfo {author} {\bibfnamefont {D.}~\bibnamefont
  {Schuricht}},\ }\href {\doibase 10.1103/PhysRevB.87.195104} {\bibfield
  {journal} {\bibinfo  {journal} {Phys. Rev. B}\ }\textbf {\bibinfo {volume}
  {87}},\ \bibinfo {pages} {195104} (\bibinfo {year} {2013})}\BibitemShut
  {NoStop}%
\bibitem [{\citenamefont {Andraschko}\ and\ \citenamefont
  {Sirker}(2014)}]{Andraschko2014}%
  \BibitemOpen
  \bibfield  {author} {\bibinfo {author} {\bibfnamefont {F.}~\bibnamefont
  {Andraschko}}\ and\ \bibinfo {author} {\bibfnamefont {J.}~\bibnamefont
  {Sirker}},\ }\href {\doibase 10.1103/PhysRevB.89.125120} {\bibfield
  {journal} {\bibinfo  {journal} {Phys. Rev. B}\ }\textbf {\bibinfo {volume}
  {89}},\ \bibinfo {pages} {125120} (\bibinfo {year} {2014})}\BibitemShut
  {NoStop}%
\bibitem [{\citenamefont {Vajna}\ and\ \citenamefont
  {D\'ora}(2014)}]{Vajna2014}%
  \BibitemOpen
  \bibfield  {author} {\bibinfo {author} {\bibfnamefont {S.}~\bibnamefont
  {Vajna}}\ and\ \bibinfo {author} {\bibfnamefont {B.}~\bibnamefont {D\'ora}},\
  }\href {\doibase 10.1103/PhysRevB.89.161105} {\bibfield  {journal} {\bibinfo
  {journal} {Phys. Rev. B}\ }\textbf {\bibinfo {volume} {89}},\ \bibinfo
  {pages} {161105} (\bibinfo {year} {2014})}\BibitemShut {NoStop}%
\bibitem [{\citenamefont {Vajna}\ and\ \citenamefont
  {D\'ora}(2015)}]{Vajna2015}%
  \BibitemOpen
  \bibfield  {author} {\bibinfo {author} {\bibfnamefont {S.}~\bibnamefont
  {Vajna}}\ and\ \bibinfo {author} {\bibfnamefont {B.}~\bibnamefont {D\'ora}},\
  }\href {\doibase 10.1103/PhysRevB.91.155127} {\bibfield  {journal} {\bibinfo
  {journal} {Phys. Rev. B}\ }\textbf {\bibinfo {volume} {91}},\ \bibinfo
  {pages} {155127} (\bibinfo {year} {2015})}\BibitemShut {NoStop}%
\bibitem [{\citenamefont {Budich}\ and\ \citenamefont
  {Heyl}(2016)}]{Budich2016}%
  \BibitemOpen
  \bibfield  {author} {\bibinfo {author} {\bibfnamefont {J.~C.}\ \bibnamefont
  {Budich}}\ and\ \bibinfo {author} {\bibfnamefont {M.}~\bibnamefont {Heyl}},\
  }\href {\doibase 10.1103/PhysRevB.93.085416} {\bibfield  {journal} {\bibinfo
  {journal} {Phys. Rev. B}\ }\textbf {\bibinfo {volume} {93}},\ \bibinfo
  {pages} {085416} (\bibinfo {year} {2016})}\BibitemShut {NoStop}%
\bibitem [{\citenamefont {Zvyagin}(2016)}]{Zvyagin2016}%
  \BibitemOpen
  \bibfield  {author} {\bibinfo {author} {\bibfnamefont {A.~A.}\ \bibnamefont
  {Zvyagin}},\ }\href {\doibase 10.1063/1.4969869} {\bibfield  {journal}
  {\bibinfo  {journal} {J. Low Temp. Phys.}\ }\textbf {\bibinfo {volume}
  {42}},\ \bibinfo {pages} {971} (\bibinfo {year} {2016})}\BibitemShut
  {NoStop}%
\bibitem [{\citenamefont {Halimeh}\ and\ \citenamefont
  {Zauner-Stauber}(2017)}]{Halimeh2017}%
  \BibitemOpen
  \bibfield  {author} {\bibinfo {author} {\bibfnamefont {J.~C.}\ \bibnamefont
  {Halimeh}}\ and\ \bibinfo {author} {\bibfnamefont {V.}~\bibnamefont
  {Zauner-Stauber}},\ }\href {\doibase 10.1103/PhysRevB.96.134427} {\bibfield
  {journal} {\bibinfo  {journal} {Phys. Rev. B}\ }\textbf {\bibinfo {volume}
  {96}},\ \bibinfo {pages} {134427} (\bibinfo {year} {2017})}\BibitemShut
  {NoStop}%
\bibitem [{\citenamefont {Heyl}(2018)}]{Heyl:2018}%
  \BibitemOpen
  \bibfield  {author} {\bibinfo {author} {\bibfnamefont {M.}~\bibnamefont
  {Heyl}},\ }\href {\doibase doi.org/10.1088/1361-6633/aaaf9a} {\bibfield
  {journal} {\bibinfo  {journal} {Rep. Prog. Phys.}\ }\textbf {\bibinfo
  {volume} {81}},\ \bibinfo {pages} {054001} (\bibinfo {year}
  {2018})}\BibitemShut {NoStop}%
\bibitem [{\citenamefont {Jurcevic}\ \emph {et~al.}(2017)\citenamefont
  {Jurcevic}, \citenamefont {Shen}, \citenamefont {Hauke}, \citenamefont
  {Maier}, \citenamefont {Brydges}, \citenamefont {Hempel}, \citenamefont
  {Lanyon}, \citenamefont {Heyl}, \citenamefont {Blatt},\ and\ \citenamefont
  {Roos}}]{Jurcevic2017}%
  \BibitemOpen
  \bibfield  {author} {\bibinfo {author} {\bibfnamefont {P.}~\bibnamefont
  {Jurcevic}}, \bibinfo {author} {\bibfnamefont {H.}~\bibnamefont {Shen}},
  \bibinfo {author} {\bibfnamefont {P.}~\bibnamefont {Hauke}}, \bibinfo
  {author} {\bibfnamefont {C.}~\bibnamefont {Maier}}, \bibinfo {author}
  {\bibfnamefont {T.}~\bibnamefont {Brydges}}, \bibinfo {author} {\bibfnamefont
  {C.}~\bibnamefont {Hempel}}, \bibinfo {author} {\bibfnamefont {B.~P.}\
  \bibnamefont {Lanyon}}, \bibinfo {author} {\bibfnamefont {M.}~\bibnamefont
  {Heyl}}, \bibinfo {author} {\bibfnamefont {R.}~\bibnamefont {Blatt}}, \ and\
  \bibinfo {author} {\bibfnamefont {C.~F.}\ \bibnamefont {Roos}},\ }\href
  {\doibase 10.1103/PhysRevLett.119.080501} {\bibfield  {journal} {\bibinfo
  {journal} {Phys. Rev. Lett.}\ }\textbf {\bibinfo {volume} {119}},\ \bibinfo
  {pages} {080501} (\bibinfo {year} {2017})}\BibitemShut {NoStop}%
\bibitem [{\citenamefont {Zhang}\ \emph {et~al.}(2017)\citenamefont {Zhang},
  \citenamefont {Pagano}, \citenamefont {Hess}, \citenamefont {Kyprianidis},
  \citenamefont {Becker}, \citenamefont {Kaplan}, \citenamefont {Gorshkov},
  \citenamefont {Gong},\ and\ \citenamefont {Monroe}}]{Zhang2017}%
  \BibitemOpen
  \bibfield  {author} {\bibinfo {author} {\bibfnamefont {J.}~\bibnamefont
  {Zhang}}, \bibinfo {author} {\bibfnamefont {G.}~\bibnamefont {Pagano}},
  \bibinfo {author} {\bibfnamefont {P.~W.}\ \bibnamefont {Hess}}, \bibinfo
  {author} {\bibfnamefont {A.}~\bibnamefont {Kyprianidis}}, \bibinfo {author}
  {\bibfnamefont {P.}~\bibnamefont {Becker}}, \bibinfo {author} {\bibfnamefont
  {H.}~\bibnamefont {Kaplan}}, \bibinfo {author} {\bibfnamefont {A.~V.}\
  \bibnamefont {Gorshkov}}, \bibinfo {author} {\bibfnamefont {Z.-X.}\
  \bibnamefont {Gong}}, \ and\ \bibinfo {author} {\bibfnamefont
  {C.}~\bibnamefont {Monroe}},\ }\href {https://doi.org/10.1038/nature24654}
  {\bibfield  {journal} {\bibinfo  {journal} {Nature}\ }\textbf {\bibinfo
  {volume} {551}},\ \bibinfo {pages} {601} (\bibinfo {year}
  {2017})}\BibitemShut {NoStop}%
\bibitem [{\citenamefont {Fläschner}\ \emph {et~al.}(2018)\citenamefont
  {Fläschner}, \citenamefont {Vogel}, \citenamefont {Tarnowski}, \citenamefont
  {Rem}, \citenamefont {Lühmann}, \citenamefont {Heyl}, \citenamefont
  {Budich}, \citenamefont {Mathey}, \citenamefont {Sengstock},\ and\
  \citenamefont {Weitenberg}}]{Flaeschner2018}%
  \BibitemOpen
  \bibfield  {author} {\bibinfo {author} {\bibfnamefont {N.}~\bibnamefont
  {Fläschner}}, \bibinfo {author} {\bibfnamefont {D.}~\bibnamefont {Vogel}},
  \bibinfo {author} {\bibfnamefont {M.}~\bibnamefont {Tarnowski}}, \bibinfo
  {author} {\bibfnamefont {B.~S.}\ \bibnamefont {Rem}}, \bibinfo {author}
  {\bibfnamefont {D.-S.}\ \bibnamefont {Lühmann}}, \bibinfo {author}
  {\bibfnamefont {M.}~\bibnamefont {Heyl}}, \bibinfo {author} {\bibfnamefont
  {J.~C.}\ \bibnamefont {Budich}}, \bibinfo {author} {\bibfnamefont
  {L.}~\bibnamefont {Mathey}}, \bibinfo {author} {\bibfnamefont
  {K.}~\bibnamefont {Sengstock}}, \ and\ \bibinfo {author} {\bibfnamefont
  {C.}~\bibnamefont {Weitenberg}},\ }\href
  {https://doi.org/10.1038/s41567-017-0013-8} {\bibfield  {journal} {\bibinfo
  {journal} {Nat. Phys.}\ }\textbf {\bibinfo {volume} {14}},\ \bibinfo {pages}
  {265} (\bibinfo {year} {2018})}\BibitemShut {NoStop}%
\bibitem [{\citenamefont {Guo}\ \emph {et~al.}(2019)\citenamefont {Guo},
  \citenamefont {Yang}, \citenamefont {Zeng}, \citenamefont {Peng},
  \citenamefont {Li}, \citenamefont {Deng}, \citenamefont {Jin}, \citenamefont
  {Chen}, \citenamefont {Zheng},\ and\ \citenamefont {Fan}}]{Guo2019}%
  \BibitemOpen
  \bibfield  {author} {\bibinfo {author} {\bibfnamefont {X.-Y.}\ \bibnamefont
  {Guo}}, \bibinfo {author} {\bibfnamefont {C.}~\bibnamefont {Yang}}, \bibinfo
  {author} {\bibfnamefont {Y.}~\bibnamefont {Zeng}}, \bibinfo {author}
  {\bibfnamefont {Y.}~\bibnamefont {Peng}}, \bibinfo {author} {\bibfnamefont
  {H.-K.}\ \bibnamefont {Li}}, \bibinfo {author} {\bibfnamefont
  {H.}~\bibnamefont {Deng}}, \bibinfo {author} {\bibfnamefont {Y.-R.}\
  \bibnamefont {Jin}}, \bibinfo {author} {\bibfnamefont {S.}~\bibnamefont
  {Chen}}, \bibinfo {author} {\bibfnamefont {D.}~\bibnamefont {Zheng}}, \ and\
  \bibinfo {author} {\bibfnamefont {H.}~\bibnamefont {Fan}},\ }\href {\doibase
  10.1103/PhysRevApplied.11.044080} {\bibfield  {journal} {\bibinfo  {journal}
  {Phys. Rev. Applied}\ }\textbf {\bibinfo {volume} {11}},\ \bibinfo {pages}
  {044080} (\bibinfo {year} {2019})}\BibitemShut {NoStop}%
\bibitem [{\citenamefont {Tian}\ \emph {et~al.}(2019)\citenamefont {Tian},
  \citenamefont {Ke}, \citenamefont {Zhang}, \citenamefont {Lin}, \citenamefont
  {Shi}, \citenamefont {Huang}, \citenamefont {Lee},\ and\ \citenamefont
  {Du}}]{Tian2019}%
  \BibitemOpen
  \bibfield  {author} {\bibinfo {author} {\bibfnamefont {T.}~\bibnamefont
  {Tian}}, \bibinfo {author} {\bibfnamefont {Y.}~\bibnamefont {Ke}}, \bibinfo
  {author} {\bibfnamefont {L.}~\bibnamefont {Zhang}}, \bibinfo {author}
  {\bibfnamefont {S.}~\bibnamefont {Lin}}, \bibinfo {author} {\bibfnamefont
  {Z.}~\bibnamefont {Shi}}, \bibinfo {author} {\bibfnamefont {P.}~\bibnamefont
  {Huang}}, \bibinfo {author} {\bibfnamefont {C.}~\bibnamefont {Lee}}, \ and\
  \bibinfo {author} {\bibfnamefont {J.}~\bibnamefont {Du}},\ }\href {\doibase
  10.1103/PhysRevB.100.024310} {\bibfield  {journal} {\bibinfo  {journal}
  {Phys. Rev. B}\ }\textbf {\bibinfo {volume} {100}},\ \bibinfo {pages}
  {024310} (\bibinfo {year} {2019})}\BibitemShut {NoStop}%
\bibitem [{\citenamefont {Wang}\ \emph {et~al.}(2019)\citenamefont {Wang},
  \citenamefont {Qiu}, \citenamefont {Xiao}, \citenamefont {Zhan},
  \citenamefont {Bian}, \citenamefont {Yi},\ and\ \citenamefont
  {Xue}}]{Wang2019}%
  \BibitemOpen
  \bibfield  {author} {\bibinfo {author} {\bibfnamefont {K.}~\bibnamefont
  {Wang}}, \bibinfo {author} {\bibfnamefont {X.}~\bibnamefont {Qiu}}, \bibinfo
  {author} {\bibfnamefont {L.}~\bibnamefont {Xiao}}, \bibinfo {author}
  {\bibfnamefont {X.}~\bibnamefont {Zhan}}, \bibinfo {author} {\bibfnamefont
  {Z.}~\bibnamefont {Bian}}, \bibinfo {author} {\bibfnamefont {W.}~\bibnamefont
  {Yi}}, \ and\ \bibinfo {author} {\bibfnamefont {P.}~\bibnamefont {Xue}},\
  }\href {\doibase 10.1103/PhysRevLett.122.020501} {\bibfield  {journal}
  {\bibinfo  {journal} {Phys. Rev. Lett.}\ }\textbf {\bibinfo {volume} {122}},\
  \bibinfo {pages} {020501} (\bibinfo {year} {2019})}\BibitemShut {NoStop}%
\bibitem [{\citenamefont {Tian}\ \emph {et~al.}(2020)\citenamefont {Tian},
  \citenamefont {Yang}, \citenamefont {Qiu}, \citenamefont {Liang},
  \citenamefont {Yang}, \citenamefont {Xu},\ and\ \citenamefont
  {Duan}}]{Tian2020}%
  \BibitemOpen
  \bibfield  {author} {\bibinfo {author} {\bibfnamefont {T.}~\bibnamefont
  {Tian}}, \bibinfo {author} {\bibfnamefont {H.-X.}\ \bibnamefont {Yang}},
  \bibinfo {author} {\bibfnamefont {L.-Y.}\ \bibnamefont {Qiu}}, \bibinfo
  {author} {\bibfnamefont {H.-Y.}\ \bibnamefont {Liang}}, \bibinfo {author}
  {\bibfnamefont {Y.-B.}\ \bibnamefont {Yang}}, \bibinfo {author}
  {\bibfnamefont {Y.}~\bibnamefont {Xu}}, \ and\ \bibinfo {author}
  {\bibfnamefont {L.-M.}\ \bibnamefont {Duan}},\ }\href {\doibase
  10.1103/PhysRevLett.124.043001} {\bibfield  {journal} {\bibinfo  {journal}
  {Phys. Rev. Lett.}\ }\textbf {\bibinfo {volume} {124}},\ \bibinfo {pages}
  {043001} (\bibinfo {year} {2020})}\BibitemShut {NoStop}%
\bibitem [{\citenamefont {Xu}\ \emph {et~al.}(2020)\citenamefont {Xu},
  \citenamefont {Sun}, \citenamefont {Liu}, \citenamefont {Zhang},
  \citenamefont {Li}, \citenamefont {Dong}, \citenamefont {Ren}, \citenamefont
  {Zhang}, \citenamefont {Nori}, \citenamefont {Zheng}, \citenamefont {Fan},\
  and\ \citenamefont {Wang}}]{Xu2020}%
  \BibitemOpen
  \bibfield  {author} {\bibinfo {author} {\bibfnamefont {K.}~\bibnamefont
  {Xu}}, \bibinfo {author} {\bibfnamefont {Z.-H.}\ \bibnamefont {Sun}},
  \bibinfo {author} {\bibfnamefont {W.}~\bibnamefont {Liu}}, \bibinfo {author}
  {\bibfnamefont {Y.-R.}\ \bibnamefont {Zhang}}, \bibinfo {author}
  {\bibfnamefont {H.}~\bibnamefont {Li}}, \bibinfo {author} {\bibfnamefont
  {H.}~\bibnamefont {Dong}}, \bibinfo {author} {\bibfnamefont {W.}~\bibnamefont
  {Ren}}, \bibinfo {author} {\bibfnamefont {P.}~\bibnamefont {Zhang}}, \bibinfo
  {author} {\bibfnamefont {F.}~\bibnamefont {Nori}}, \bibinfo {author}
  {\bibfnamefont {D.}~\bibnamefont {Zheng}}, \bibinfo {author} {\bibfnamefont
  {H.}~\bibnamefont {Fan}}, \ and\ \bibinfo {author} {\bibfnamefont
  {H.}~\bibnamefont {Wang}},\ }\href {\doibase 10.1126/sciadv.aba4935}
  {\bibfield  {journal} {\bibinfo  {journal} {Science Adv.}\ }\textbf {\bibinfo
  {volume} {6}},\ \bibinfo {pages} {eaba4935} (\bibinfo {year}
  {2020})}\BibitemShut {NoStop}%
\bibitem [{\citenamefont {Li}\ and\ \citenamefont {Haldane}(2008)}]{Li2007}%
  \BibitemOpen
  \bibfield  {author} {\bibinfo {author} {\bibfnamefont {H.}~\bibnamefont
  {Li}}\ and\ \bibinfo {author} {\bibfnamefont {F.~D.~M.}\ \bibnamefont
  {Haldane}},\ }\href {\doibase 10.1103/PhysRevLett.101.010504} {\bibfield
  {journal} {\bibinfo  {journal} {Phys. Rev. Lett.}\ }\textbf {\bibinfo
  {volume} {101}},\ \bibinfo {pages} {010504} (\bibinfo {year}
  {2008})}\BibitemShut {NoStop}%
\bibitem [{\citenamefont {Fidkowski}(2010)}]{Fidkowski2010}%
  \BibitemOpen
  \bibfield  {author} {\bibinfo {author} {\bibfnamefont {L.}~\bibnamefont
  {Fidkowski}},\ }\href {\doibase 10.1103/PhysRevLett.104.130502} {\bibfield
  {journal} {\bibinfo  {journal} {Phys. Rev. Lett.}\ }\textbf {\bibinfo
  {volume} {104}},\ \bibinfo {pages} {130502} (\bibinfo {year}
  {2010})}\BibitemShut {NoStop}%
\bibitem [{\citenamefont {Torlai}\ \emph {et~al.}(2014)\citenamefont {Torlai},
  \citenamefont {Tagliacozzo},\ and\ \citenamefont {Chiara}}]{Torlai2014}%
  \BibitemOpen
  \bibfield  {author} {\bibinfo {author} {\bibfnamefont {G.}~\bibnamefont
  {Torlai}}, \bibinfo {author} {\bibfnamefont {L.}~\bibnamefont {Tagliacozzo}},
  \ and\ \bibinfo {author} {\bibfnamefont {G.~D.}\ \bibnamefont {Chiara}},\
  }\href {\doibase 10.1088/1742-5468/2014/06/p06001} {\bibfield  {journal}
  {\bibinfo  {journal} {Journal of Statistical Mechanics: Theory and
  Experiment}\ }\textbf {\bibinfo {volume} {2014}},\ \bibinfo {pages} {P06001}
  (\bibinfo {year} {2014})}\BibitemShut {NoStop}%
\bibitem [{\citenamefont {Canovi}\ \emph {et~al.}(2014)\citenamefont {Canovi},
  \citenamefont {Ercolessi}, \citenamefont {Naldesi}, \citenamefont {Taddia},\
  and\ \citenamefont {Vodola}}]{Canovi2014}%
  \BibitemOpen
  \bibfield  {author} {\bibinfo {author} {\bibfnamefont {E.}~\bibnamefont
  {Canovi}}, \bibinfo {author} {\bibfnamefont {E.}~\bibnamefont {Ercolessi}},
  \bibinfo {author} {\bibfnamefont {P.}~\bibnamefont {Naldesi}}, \bibinfo
  {author} {\bibfnamefont {L.}~\bibnamefont {Taddia}}, \ and\ \bibinfo {author}
  {\bibfnamefont {D.}~\bibnamefont {Vodola}},\ }\href {\doibase
  10.1103/PhysRevB.89.104303} {\bibfield  {journal} {\bibinfo  {journal} {Phys.
  Rev. B}\ }\textbf {\bibinfo {volume} {89}},\ \bibinfo {pages} {104303}
  (\bibinfo {year} {2014})}\BibitemShut {NoStop}%
\bibitem [{\citenamefont {Jafari}\ \emph {et~al.}(2019)\citenamefont {Jafari},
  \citenamefont {Johannesson}, \citenamefont {Langari},\ and\ \citenamefont
  {Martin-Delgado}}]{Jafari2019}%
  \BibitemOpen
  \bibfield  {author} {\bibinfo {author} {\bibfnamefont {R.}~\bibnamefont
  {Jafari}}, \bibinfo {author} {\bibfnamefont {H.}~\bibnamefont {Johannesson}},
  \bibinfo {author} {\bibfnamefont {A.}~\bibnamefont {Langari}}, \ and\
  \bibinfo {author} {\bibfnamefont {M.~A.}\ \bibnamefont {Martin-Delgado}},\
  }\href {\doibase 10.1103/PhysRevB.99.054302} {\bibfield  {journal} {\bibinfo
  {journal} {Phys. Rev. B}\ }\textbf {\bibinfo {volume} {99}},\ \bibinfo
  {pages} {054302} (\bibinfo {year} {2019})}\BibitemShut {NoStop}%
\bibitem [{\citenamefont {De~Nicola}\ \emph {et~al.}(2021)\citenamefont
  {De~Nicola}, \citenamefont {Michailidis},\ and\ \citenamefont
  {Serbyn}}]{DeNicola2021}%
  \BibitemOpen
  \bibfield  {author} {\bibinfo {author} {\bibfnamefont {S.}~\bibnamefont
  {De~Nicola}}, \bibinfo {author} {\bibfnamefont {A.~A.}\ \bibnamefont
  {Michailidis}}, \ and\ \bibinfo {author} {\bibfnamefont {M.}~\bibnamefont
  {Serbyn}},\ }\href {\doibase 10.1103/PhysRevLett.126.040602} {\bibfield
  {journal} {\bibinfo  {journal} {Phys. Rev. Lett.}\ }\textbf {\bibinfo
  {volume} {126}},\ \bibinfo {pages} {040602} (\bibinfo {year}
  {2021})}\BibitemShut {NoStop}%
\bibitem [{\citenamefont {Gong}\ and\ \citenamefont {Ueda}(2018)}]{gong2018}%
  \BibitemOpen
  \bibfield  {author} {\bibinfo {author} {\bibfnamefont {Z.}~\bibnamefont
  {Gong}}\ and\ \bibinfo {author} {\bibfnamefont {M.}~\bibnamefont {Ueda}},\
  }\href {\doibase 10.1103/PhysRevLett.121.250601} {\bibfield  {journal}
  {\bibinfo  {journal} {Phys. Rev. Lett.}\ }\textbf {\bibinfo {volume} {121}},\
  \bibinfo {pages} {250601} (\bibinfo {year} {2018})}\BibitemShut {NoStop}%
\bibitem [{\citenamefont {Gong}\ \emph {et~al.}(2019)\citenamefont {Gong},
  \citenamefont {Kura}, \citenamefont {Sato},\ and\ \citenamefont
  {Ueda}}]{gong2019}%
  \BibitemOpen
  \bibfield  {author} {\bibinfo {author} {\bibfnamefont {Z.}~\bibnamefont
  {Gong}}, \bibinfo {author} {\bibfnamefont {N.}~\bibnamefont {Kura}}, \bibinfo
  {author} {\bibfnamefont {M.}~\bibnamefont {Sato}}, \ and\ \bibinfo {author}
  {\bibfnamefont {M.}~\bibnamefont {Ueda}},\ }\href@noop {} {\  (\bibinfo
  {year} {2019})},\ \Eprint {http://arxiv.org/abs/1904.12464} {arXiv:1904.12464
  [quant-ph]} \BibitemShut {NoStop}%
\bibitem [{\citenamefont {Lu}\ and\ \citenamefont {Yu}(2019)}]{Lu2019}%
  \BibitemOpen
  \bibfield  {author} {\bibinfo {author} {\bibfnamefont {S.}~\bibnamefont
  {Lu}}\ and\ \bibinfo {author} {\bibfnamefont {J.}~\bibnamefont {Yu}},\ }\href
  {\doibase 10.1103/PhysRevA.99.033621} {\bibfield  {journal} {\bibinfo
  {journal} {Phys. Rev. A}\ }\textbf {\bibinfo {volume} {99}},\ \bibinfo
  {pages} {033621} (\bibinfo {year} {2019})}\BibitemShut {NoStop}%
\bibitem [{\citenamefont {Pastori}\ \emph {et~al.}(2020)\citenamefont
  {Pastori}, \citenamefont {Barbarino},\ and\ \citenamefont
  {Budich}}]{Pastori2020}%
  \BibitemOpen
  \bibfield  {author} {\bibinfo {author} {\bibfnamefont {L.}~\bibnamefont
  {Pastori}}, \bibinfo {author} {\bibfnamefont {S.}~\bibnamefont {Barbarino}},
  \ and\ \bibinfo {author} {\bibfnamefont {J.~C.}\ \bibnamefont {Budich}},\
  }\href {\doibase 10.1103/PhysRevResearch.2.033259} {\bibfield  {journal}
  {\bibinfo  {journal} {Phys. Rev. Research}\ }\textbf {\bibinfo {volume}
  {2}},\ \bibinfo {pages} {033259} (\bibinfo {year} {2020})}\BibitemShut
  {NoStop}%
\bibitem [{\citenamefont {Surace}\ \emph {et~al.}(2020)\citenamefont {Surace},
  \citenamefont {Tagliacozzo},\ and\ \citenamefont {Tonni}}]{Surace2020}%
  \BibitemOpen
  \bibfield  {author} {\bibinfo {author} {\bibfnamefont {J.}~\bibnamefont
  {Surace}}, \bibinfo {author} {\bibfnamefont {L.}~\bibnamefont {Tagliacozzo}},
  \ and\ \bibinfo {author} {\bibfnamefont {E.}~\bibnamefont {Tonni}},\ }\href
  {\doibase 10.1103/PhysRevB.101.241107} {\bibfield  {journal} {\bibinfo
  {journal} {Phys. Rev. B}\ }\textbf {\bibinfo {volume} {101}},\ \bibinfo
  {pages} {241107} (\bibinfo {year} {2020})}\BibitemShut {NoStop}%
\bibitem [{\citenamefont {Ekert}\ and\ \citenamefont
  {Knight}(1995)}]{Ekert1995}%
  \BibitemOpen
  \bibfield  {author} {\bibinfo {author} {\bibfnamefont {A.}~\bibnamefont
  {Ekert}}\ and\ \bibinfo {author} {\bibfnamefont {P.~L.}\ \bibnamefont
  {Knight}},\ }\href {\doibase 10.1119/1.17904} {\bibfield  {journal} {\bibinfo
   {journal} {American Journal of Physics}\ }\textbf {\bibinfo {volume} {63}},\
  \bibinfo {pages} {415} (\bibinfo {year} {1995})},\ \Eprint
  {http://arxiv.org/abs/https://doi.org/10.1119/1.17904}
  {https://doi.org/10.1119/1.17904} \BibitemShut {NoStop}%
\bibitem [{\citenamefont {Peschel}(2003)}]{Peschel_2003}%
  \BibitemOpen
  \bibfield  {author} {\bibinfo {author} {\bibfnamefont {I.}~\bibnamefont
  {Peschel}},\ }\href {\doibase 10.1088/0305-4470/36/14/101} {\bibfield
  {journal} {\bibinfo  {journal} {Journal of Physics A: Mathematical and
  General}\ }\textbf {\bibinfo {volume} {36}},\ \bibinfo {pages} {L205}
  (\bibinfo {year} {2003})}\BibitemShut {NoStop}%
\bibitem [{\citenamefont {Peschel}\ and\ \citenamefont
  {Eisler}(2009)}]{Peschel_2009}%
  \BibitemOpen
  \bibfield  {author} {\bibinfo {author} {\bibfnamefont {I.}~\bibnamefont
  {Peschel}}\ and\ \bibinfo {author} {\bibfnamefont {V.}~\bibnamefont
  {Eisler}},\ }\href {\doibase 10.1088/1751-8113/42/50/504003} {\bibfield
  {journal} {\bibinfo  {journal} {Journal of Physics A: Mathematical and
  Theoretical}\ }\textbf {\bibinfo {volume} {42}},\ \bibinfo {pages} {504003}
  (\bibinfo {year} {2009})}\BibitemShut {NoStop}%
\bibitem [{\citenamefont {Najafi}\ \emph {et~al.}(2019)\citenamefont {Najafi},
  \citenamefont {Rajabpour},\ and\ \citenamefont {Viti}}]{Najafi2019}%
  \BibitemOpen
  \bibfield  {author} {\bibinfo {author} {\bibfnamefont {K.}~\bibnamefont
  {Najafi}}, \bibinfo {author} {\bibfnamefont {M.~A.}\ \bibnamefont
  {Rajabpour}}, \ and\ \bibinfo {author} {\bibfnamefont {J.}~\bibnamefont
  {Viti}},\ }\href {\doibase 10.1088/1742-5468/ab3413} {\bibfield  {journal}
  {\bibinfo  {journal} {J. Stat. Mech.}\ }\textbf {\bibinfo {volume} {2019}},\
  \bibinfo {pages} {083102} (\bibinfo {year} {2019})}\BibitemShut {NoStop}%
\bibitem [{\citenamefont {Zamani}\ \emph {et~al.}(2020)\citenamefont {Zamani},
  \citenamefont {Jafari},\ and\ \citenamefont {Langari}}]{zamani2020}%
  \BibitemOpen
  \bibfield  {author} {\bibinfo {author} {\bibfnamefont {S.}~\bibnamefont
  {Zamani}}, \bibinfo {author} {\bibfnamefont {R.}~\bibnamefont {Jafari}}, \
  and\ \bibinfo {author} {\bibfnamefont {A.}~\bibnamefont {Langari}},\ }\href
  {\doibase 10.1103/PhysRevB.102.144306} {\bibfield  {journal} {\bibinfo
  {journal} {Phys. Rev. B}\ }\textbf {\bibinfo {volume} {102}},\ \bibinfo
  {pages} {144306} (\bibinfo {year} {2020})}\BibitemShut {NoStop}%
\bibitem [{\citenamefont {Schmitt}\ and\ \citenamefont
  {Kehrein}(2015)}]{Schmitt2015}%
  \BibitemOpen
  \bibfield  {author} {\bibinfo {author} {\bibfnamefont {M.}~\bibnamefont
  {Schmitt}}\ and\ \bibinfo {author} {\bibfnamefont {S.}~\bibnamefont
  {Kehrein}},\ }\href {\doibase 10.1103/PhysRevB.92.075114} {\bibfield
  {journal} {\bibinfo  {journal} {Phys. Rev. B}\ }\textbf {\bibinfo {volume}
  {92}},\ \bibinfo {pages} {075114} (\bibinfo {year} {2015})}\BibitemShut
  {NoStop}%
\bibitem [{\citenamefont {Altland}\ and\ \citenamefont
  {Zirnbauer}(1997)}]{Altland1997}%
  \BibitemOpen
  \bibfield  {author} {\bibinfo {author} {\bibfnamefont {A.}~\bibnamefont
  {Altland}}\ and\ \bibinfo {author} {\bibfnamefont {M.~R.}\ \bibnamefont
  {Zirnbauer}},\ }\href {\doibase 10.1103/PhysRevB.55.1142} {\bibfield
  {journal} {\bibinfo  {journal} {Phys. Rev. B}\ }\textbf {\bibinfo {volume}
  {55}},\ \bibinfo {pages} {1142} (\bibinfo {year} {1997})}\BibitemShut
  {NoStop}%
\bibitem [{\citenamefont {Schnyder}\ \emph {et~al.}(2008)\citenamefont
  {Schnyder}, \citenamefont {Ryu}, \citenamefont {Furusaki},\ and\
  \citenamefont {Ludwig}}]{Schnyder2008}%
  \BibitemOpen
  \bibfield  {author} {\bibinfo {author} {\bibfnamefont {A.~P.}\ \bibnamefont
  {Schnyder}}, \bibinfo {author} {\bibfnamefont {S.}~\bibnamefont {Ryu}},
  \bibinfo {author} {\bibfnamefont {A.}~\bibnamefont {Furusaki}}, \ and\
  \bibinfo {author} {\bibfnamefont {A.~W.~W.}\ \bibnamefont {Ludwig}},\ }\href
  {\doibase 10.1103/PhysRevB.78.195125} {\bibfield  {journal} {\bibinfo
  {journal} {Phys. Rev. B}\ }\textbf {\bibinfo {volume} {78}},\ \bibinfo
  {pages} {195125} (\bibinfo {year} {2008})}\BibitemShut {NoStop}%
\bibitem [{\citenamefont {Bhattacharya}\ and\ \citenamefont
  {Dutta}(2017)}]{bhattacharya2017}%
  \BibitemOpen
  \bibfield  {author} {\bibinfo {author} {\bibfnamefont {U.}~\bibnamefont
  {Bhattacharya}}\ and\ \bibinfo {author} {\bibfnamefont {A.}~\bibnamefont
  {Dutta}},\ }\href {\doibase 10.1103/PhysRevB.96.014302} {\bibfield  {journal}
  {\bibinfo  {journal} {Phys. Rev. B}\ }\textbf {\bibinfo {volume} {96}},\
  \bibinfo {pages} {014302} (\bibinfo {year} {2017})}\BibitemShut {NoStop}%
\bibitem [{\citenamefont {Heyl}\ and\ \citenamefont {Budich}(2017)}]{heyl2017}%
  \BibitemOpen
  \bibfield  {author} {\bibinfo {author} {\bibfnamefont {M.}~\bibnamefont
  {Heyl}}\ and\ \bibinfo {author} {\bibfnamefont {J.~C.}\ \bibnamefont
  {Budich}},\ }\href {\doibase 10.1103/PhysRevB.96.180304} {\bibfield
  {journal} {\bibinfo  {journal} {Phys. Rev. B}\ }\textbf {\bibinfo {volume}
  {96}},\ \bibinfo {pages} {180304} (\bibinfo {year} {2017})}\BibitemShut
  {NoStop}%
\bibitem [{\citenamefont {Sedlmayr}\ \emph
  {et~al.}(2018{\natexlab{a}})\citenamefont {Sedlmayr}, \citenamefont
  {Fleischhauer},\ and\ \citenamefont {Sirker}}]{Sedlmayr2018open}%
  \BibitemOpen
  \bibfield  {author} {\bibinfo {author} {\bibfnamefont {N.}~\bibnamefont
  {Sedlmayr}}, \bibinfo {author} {\bibfnamefont {M.}~\bibnamefont
  {Fleischhauer}}, \ and\ \bibinfo {author} {\bibfnamefont {J.}~\bibnamefont
  {Sirker}},\ }\href {\doibase 10.1103/PhysRevB.97.045147} {\bibfield
  {journal} {\bibinfo  {journal} {Phys. Rev. B}\ }\textbf {\bibinfo {volume}
  {97}},\ \bibinfo {pages} {045147} (\bibinfo {year}
  {2018}{\natexlab{a}})}\BibitemShut {NoStop}%
\bibitem [{\citenamefont {Abeling}\ and\ \citenamefont
  {Kehrein}(2016)}]{abelin2016}%
  \BibitemOpen
  \bibfield  {author} {\bibinfo {author} {\bibfnamefont {N.~O.}\ \bibnamefont
  {Abeling}}\ and\ \bibinfo {author} {\bibfnamefont {S.}~\bibnamefont
  {Kehrein}},\ }\href {\doibase 10.1103/PhysRevB.93.104302} {\bibfield
  {journal} {\bibinfo  {journal} {Phys. Rev. B}\ }\textbf {\bibinfo {volume}
  {93}},\ \bibinfo {pages} {104302} (\bibinfo {year} {2016})}\BibitemShut
  {NoStop}%
\bibitem [{\citenamefont {Zhang}\ \emph {et~al.}(2014)\citenamefont {Zhang},
  \citenamefont {Sheng}, \citenamefont {Shen}, \citenamefont {Wang},\ and\
  \citenamefont {Xing}}]{Zhang_2014}%
  \BibitemOpen
  \bibfield  {author} {\bibinfo {author} {\bibfnamefont {Y.~F.}\ \bibnamefont
  {Zhang}}, \bibinfo {author} {\bibfnamefont {L.}~\bibnamefont {Sheng}},
  \bibinfo {author} {\bibfnamefont {R.}~\bibnamefont {Shen}}, \bibinfo {author}
  {\bibfnamefont {R.}~\bibnamefont {Wang}}, \ and\ \bibinfo {author}
  {\bibfnamefont {D.~Y.}\ \bibnamefont {Xing}},\ }\href {\doibase
  10.1088/0953-8984/26/10/105502} {\bibfield  {journal} {\bibinfo  {journal}
  {Journal of Physics: Condensed Matter}\ }\textbf {\bibinfo {volume} {26}},\
  \bibinfo {pages} {105502} (\bibinfo {year} {2014})}\BibitemShut {NoStop}%
\bibitem [{\citenamefont {Sedlmayr}\ \emph
  {et~al.}(2018{\natexlab{b}})\citenamefont {Sedlmayr}, \citenamefont {Jaeger},
  \citenamefont {Maiti},\ and\ \citenamefont {Sirker}}]{Sedlmayr2018}%
  \BibitemOpen
  \bibfield  {author} {\bibinfo {author} {\bibfnamefont {N.}~\bibnamefont
  {Sedlmayr}}, \bibinfo {author} {\bibfnamefont {P.}~\bibnamefont {Jaeger}},
  \bibinfo {author} {\bibfnamefont {M.}~\bibnamefont {Maiti}}, \ and\ \bibinfo
  {author} {\bibfnamefont {J.}~\bibnamefont {Sirker}},\ }\href {\doibase
  10.1103/PhysRevB.97.064304} {\bibfield  {journal} {\bibinfo  {journal} {Phys.
  Rev. B}\ }\textbf {\bibinfo {volume} {97}},\ \bibinfo {pages} {064304}
  (\bibinfo {year} {2018}{\natexlab{b}})}\BibitemShut {NoStop}%
\bibitem [{\citenamefont {Kuo}\ \emph {et~al.}(2021)\citenamefont {Kuo},
  \citenamefont {Arovas}, \citenamefont {Vishveshwara},\ and\ \citenamefont
  {You}}]{kuo2021}%
  \BibitemOpen
  \bibfield  {author} {\bibinfo {author} {\bibfnamefont {W.-T.}\ \bibnamefont
  {Kuo}}, \bibinfo {author} {\bibfnamefont {D.}~\bibnamefont {Arovas}},
  \bibinfo {author} {\bibfnamefont {S.}~\bibnamefont {Vishveshwara}}, \ and\
  \bibinfo {author} {\bibfnamefont {Y.-Z.}\ \bibnamefont {You}},\ }\href@noop
  {} {\  (\bibinfo {year} {2021})},\ \Eprint {http://arxiv.org/abs/2103.08068}
  {arXiv:2103.08068 [quant-ph]} \BibitemShut {NoStop}%
\bibitem [{\citenamefont {Li}\ \emph {et~al.}(2018)\citenamefont {Li},
  \citenamefont {Chen},\ and\ \citenamefont {Fisher}}]{li2018}%
  \BibitemOpen
  \bibfield  {author} {\bibinfo {author} {\bibfnamefont {Y.}~\bibnamefont
  {Li}}, \bibinfo {author} {\bibfnamefont {X.}~\bibnamefont {Chen}}, \ and\
  \bibinfo {author} {\bibfnamefont {M.~P.~A.}\ \bibnamefont {Fisher}},\ }\href
  {\doibase 10.1103/PhysRevB.98.205136} {\bibfield  {journal} {\bibinfo
  {journal} {Phys. Rev. B}\ }\textbf {\bibinfo {volume} {98}},\ \bibinfo
  {pages} {205136} (\bibinfo {year} {2018})}\BibitemShut {NoStop}%
\bibitem [{\citenamefont {Chan}\ \emph {et~al.}(2019)\citenamefont {Chan},
  \citenamefont {Nandkishore}, \citenamefont {Pretko},\ and\ \citenamefont
  {Smith}}]{Chan2019}%
  \BibitemOpen
  \bibfield  {author} {\bibinfo {author} {\bibfnamefont {A.}~\bibnamefont
  {Chan}}, \bibinfo {author} {\bibfnamefont {R.~M.}\ \bibnamefont
  {Nandkishore}}, \bibinfo {author} {\bibfnamefont {M.}~\bibnamefont {Pretko}},
  \ and\ \bibinfo {author} {\bibfnamefont {G.}~\bibnamefont {Smith}},\ }\href
  {\doibase 10.1103/PhysRevB.99.224307} {\bibfield  {journal} {\bibinfo
  {journal} {Phys. Rev. B}\ }\textbf {\bibinfo {volume} {99}},\ \bibinfo
  {pages} {224307} (\bibinfo {year} {2019})}\BibitemShut {NoStop}%
\bibitem [{\citenamefont {Skinner}\ \emph {et~al.}(2019)\citenamefont
  {Skinner}, \citenamefont {Ruhman},\ and\ \citenamefont
  {Nahum}}]{skinner2019}%
  \BibitemOpen
  \bibfield  {author} {\bibinfo {author} {\bibfnamefont {B.}~\bibnamefont
  {Skinner}}, \bibinfo {author} {\bibfnamefont {J.}~\bibnamefont {Ruhman}}, \
  and\ \bibinfo {author} {\bibfnamefont {A.}~\bibnamefont {Nahum}},\ }\href
  {\doibase 10.1103/PhysRevX.9.031009} {\bibfield  {journal} {\bibinfo
  {journal} {Phys. Rev. X}\ }\textbf {\bibinfo {volume} {9}},\ \bibinfo {pages}
  {031009} (\bibinfo {year} {2019})}\BibitemShut {NoStop}%
\end{thebibliography}%

\bibliographystyle{apsrev4-1}

\pagebreak
\widetext
\begin{center}
\Large SUPPLEMENTAL INFORMATION to ``Entanglement Echo and dynamical entanglement transitions''
\end{center}

\section{Derivation of the dynamical correlation matrix}
The free fermion entanglement properties can be obtained from  the correlation matrix of a subsystem. Here we provide the detailed calculation of the dynamical correlation matrix for two-band Fermi systems employed in the main text, with number of additional results for completeness. As a starting point, we assume the system is described by the two-band Hamiltonian of Eq.~(4) of the main text,

\begin{equation}\label{eq:twoband}
H^{i/f}=\sum_{l,m}\hat{\bs{c}}_l^\dagger\left[\vec{d}^{i/f}_{lm}\cdot \bs{\sigma}\right]\hat{\bs{c}}_m,    
\end{equation}
where $\vec d_{lm}^{i/f}$ is a position-dependent four-component set of matrices parametrizing the Hamiltonian, $\bs \sigma = \begin{pmatrix}
\sigma_0, \sigma_x, \sigma_y, \sigma_z
\end{pmatrix}$ is a vector of Pauli matrices with $\sigma_0$ the $2\times 2$ unit matrix, $\hat c_l = (c_{l,\uparrow},\ c_{l,\downarrow})$ is an annihilation spinor at position $l$, and the superscript $i/f$ refers to pre-/post-quench form of the Hamiltonian.

\subsection{General case}

In the following, we will use $\hat c, \psi$ and $E$ to refer to annihilation operators, wavefunction and energies of the post-quench Hamiltonian, and equivalently $\hat a$, $\phi$ and $\epsilon$ for the pre-quench Hamiltonian. The time-dependent correlation matrix defined as
\begin{equation}
	\mathcal C_{lm}^{\sigma\sigma'}(t)=\langle\hat{c}^\dagger_{l\sigma}(t)\hat{c}_{m\sigma'}(t)\rangle,\label{eq:cmatrix_kspace_general}
\end{equation}
where the expectation value is taken in the ground state or a finite temperature state of the pre-quench Hamiltonian. Rewriting the operators in the eigenbasis of the post-quench Hamiltonian $c_{m\sigma} = \sum_E \psi_{m\sigma}^E c_E$ where their time-dependence becomes simple yields
\begin{equation}
	\mathcal C_{lm}^{\sigma\sigma'}(t) =  \sum_{E,E'}(e^{-iEt}\psi_{l\sigma}^E )^* e^{-iE't}\psi_{m\sigma'}^{E'}\langle\hat{c}^\dagger_{E}\hat{c}_{E'}\rangle
\end{equation}
In order to calculate the expectation value, we express the post-quench operators in the pre-quench eigenbasis  $\hat c_E = \sum_\epsilon \langle \psi^E||\phi^\epsilon \rangle \hat a_\epsilon$ to obtain
\begin{align}
	\mathcal C_{lm}^{\sigma\sigma'}(t) &= \sum_{E,E'}(e^{-iEt}\psi_{l\sigma}^E )^* e^{-iE't}\psi_{m\sigma'}^{E'}\sum_{\epsilon, \epsilon'} \langle\phi^\epsilon|\psi^E\rangle\langle\psi^{E'}|\phi^{\epsilon'}\rangle \langle\hat{a}^\dagger_{\epsilon}\hat{a}_{\epsilon'}\rangle
	= \langle\Psi_{m\sigma'}(t)| M |\Psi_{l\sigma}(t)\rangle\TG\label{eq:cmat_general}
\end{align}
where we have defined
\begin{align}
	|\Psi_{n\sigma}(t)\rangle &= \sum_{E}(e^{-iEt}\psi_{n\sigma}^E )^*\ket{\psi^E}\\
	M &= \sum_{\epsilon} n_F(\epsilon) \ket{\phi^{\epsilon}} \bra{\phi^\epsilon}.
\end{align}
In the above, $n_F(\epsilon) = \left[e^{\epsilon/(k_BT)} + 1\right]^{-1}$ is the Fermi-Dirac distribution.

\subsection{Translation invariant case}

In the translation invariant case, we can write the dynamical correlation matrix completely analytically. To streamline notation, we present the derivation for the 1d case and state the result for higher dimensions in the end. If translation invariance is preserved throughout the quench, it is convenient to begin by moving to $k$ space:
\begin{equation}
	\mathcal{C}_{lm}^{\sigma\sigma'}(t)=\langle\hat{c}^\dagger_{l\sigma}(t)\hat{c}_{m\sigma'}(t)\rangle=\frac{1}{L}\sum_{k}e^{-i(l-m)k} \langle\hat{c}^\dagger_{k\sigma}(t)\hat{c}_{k\sigma'}(t)\rangle,
\end{equation}
where, as before, the expectation value is evaluated in the ground state or a finite temperature state of the pre-quench Hamiltonian. The spatial indices $l,m$ label positions in the reduced subsystem $A$ while the whole system is assumed to have periodic boundary conditions. 

We transform the operators to the basis of the post-quench Hamiltonian energy eigenstates as per $\hat{c}_{k\sigma} = u_{k\sigma+}\hat c_{\omega+} + u_{k\sigma-}\hat c_{\omega-}$ where $u_{k\sigma\pm}$ are the eigenfunctions of the post-quench Hamiltonian. Again, the post-quench operators have simple exponential time dependence, so we obtain
\begin{align}\label{eq:correlation1}
	\mathcal{C}_{lm}^{\sigma\sigma'}(t)&= \frac{1}{L}\sum_{k}e^{-i(l-m)k} \langle(u_{k\sigma+}^*\hat{c}^\dagger_{\omega+}(t)+u_{k\sigma-}^*\hat{c}^\dagger_{\omega-}(t) )(u_{k\sigma'+}\hat{c}_{\omega+}(t)+u_{k\sigma'-}\hat{c}_{\omega-}(t) ) \rangle\nonumber\\
	&=\nonumber
	\frac{1}{L}\sum_{k}e^{-ilk+imk}\left[u_{k\sigma+}^*u_{k\sigma'+} \langle\hat{c}^\dagger_{\omega+}\hat{c}_{\omega+}\rangle+u_{k\sigma-}^*u_{k\sigma'-} \langle\hat{c}^\dagger_{\omega-}\hat{c}_{\omega-}\rangle\right.\nonumber\\
	&\left.\qquad+ u_{k\sigma+}^*u_{k\sigma'-}e^{i(\omega_k^+-\omega_{k}^-)t} \langle\hat{c}^\dagger_{\omega+}\hat{c}_{\omega-}\rangle+ u_{k\sigma-}^*u_{k\sigma'+}e^{i(\omega_k^--\omega_{k}^+)t} \langle\hat{c}^\dagger_{\omega-}\hat{c}_{\omega+}\rangle \right],
\end{align}
where $\omega^\pm_k$ are the eigenvalues of the two bands of the post-quench Hamiltonian. The expectation values are to be taken over the pre-quench state, while the operators here are in the post-quench state format. To evaluate the expectation values we hence convert the operators to the eigenbasis of the pre-quench Hamiltonian
\begin{equation}\label{eq:transform}
	\hat{c}_{\omega\pm}= \langle u_{k\pm}| v_{k+}\rangle\hat{a}_{\epsilon+} + \langle u_{k\pm}| v_{k-}\rangle\hat{a}_{\epsilon-}(0),
\end{equation}
where we label the pre-quench states with $v$ and $\epsilon$ analogously to the $u$ and $\omega$ of the post-quench states. We then have
\begin{equation}
	\langle\hat{c}^\dagger_{\omega\pm}\hat{c}_{\omega\pm'}\rangle
	=  \langle v_{k,-}| u_{k,\pm}\rangle \langle u_{k\pm'}| v_{k,-}\rangle n_F(\epsilon_k^-) + \langle v_{k,+}| u_{k\pm}\rangle \langle u_{k\pm'}| v_{k,+}\rangle n_F(\omega_k^+),
\end{equation}
Hence, we have 
\begin{align*}
	\mathcal{C}_{lm}^{\sigma\sigma'}(t) = \frac{1}{L}\sum_{k}e^{-i(l-m)k}&\left[u_{k\sigma+}^*u_{k\sigma'+}  \left(|\langle u_{k,+}| v_{k,-}\rangle|^2n_F(\epsilon_k^-) + |\langle u_{k,+}| v_{k,+}\rangle|^2n_F(\epsilon_k^+)\right)\right.\\
	&+u_{k\sigma-}^*u_{k\sigma'-} \left(|\langle u_{k,-}| v_{k,-}\rangle|^2n_F(\epsilon_k^-) + |\langle u_{k,-}| v_{k,+}\rangle|^2n_F(\epsilon_k^+)\right)\\
	& u_{k\sigma+}^*u_{k\sigma'-}e^{i\Delta\omega_kt} \left(\langle v_{k,-}| u_{k,+}\rangle\langle u_{k,-}| v_{k,-}\rangle n_F(\epsilon_k^-) + \langle v_{k,+}| u_{k,+}\rangle\langle u_{k,-}| v_{k,+}\rangle n_F(\epsilon_k^+) \right)\\
	&\left.+ u_{k\sigma-}^*u_{k\sigma'+}e^{-i\Delta\omega_kt} \left(\langle v_{k,-}| u_{k,-}\rangle\langle u_{k,+}| v_{k,-}\rangle n_F(\epsilon_k^-) + \langle v_{k,+}| u_{k,-}\rangle\langle u_{k,+}| v_{k,+}\rangle n_F(\epsilon_k^+) \right)\right].
\end{align*}
where $\Delta\omega \equiv \omega_k^+ - \omega_k^-$. Now, we notice that we can regroup things in terms of the vectors
\begin{align}
	x_{k\sigma}(t) &\equiv u_{k\sigma+} e^{-i \omega^+_kt}\langle u_{k,+}| v_{k,-}\rangle + e^{-i\omega_k^-t}v_{k\sigma-} \langle u_{k,-}|  v_{k,-}\rangle\\
	y_{k\sigma}(t) &\equiv u_{k\sigma+} e^{-i \omega^+_kt}\langle u_{k,+}| v_{k,+}\rangle + e^{-i\omega_k^-t}u_{k\sigma-} \langle u_{k,-}| v_{k,+}\rangle,
\end{align}
and finally obtain
\begin{equation}
	\mathcal{C}_{lm}^{\sigma\sigma'}(t)=\frac{1}{L}\sum_{k}e^{-i(l-m)k}\left[x_{k\sigma}^*(t) x_{k\sigma'}(t)n_F(\epsilon_k^-) + y_{k\sigma}^*(t) y_{k\sigma'} (t)n_F(\epsilon_k^+)\right].
\end{equation}
The result can also be brought to a form more similar to the general case solution above by defining
\begin{equation}
	|\Psi_{n\sigma,k}(t)\rangle =  \sum_{\eta = \pm 1} e^{-ikn} \left(e^{-i\omega_{k}^\eta t} u_{k,\sigma,\eta }\right)^*\ket{ u_{k,\eta}}\qquad \qquad M_{k,k'} = \delta_{k,k'}\sum_{\eta = \pm 1} n_F(\epsilon_{k}^\eta) \ket{ v_{k,\eta}}  \bra{ v_{k',\eta}},
\end{equation}
in terms of which we have
\begin{equation}
	\mathcal C_{lm}^{\sigma\sigma'}(t) = \sum_{k,k'}\langle\Psi_{m\sigma',k}| M_{k,k'} |\Psi_{l\sigma,k'}(t)\rangle.\label{eq:cinkspace_final}
\end{equation}
The final result for 1d case can be generalized to arbitrary dimension straightforward along the lines presented above. The result for $n$ spatial dimensions is 
\begin{equation}
	\mathcal{C}_{\mathbf{l} \mathbf{m}}^{\sigma\sigma'}(t)=\frac{1}{\Omega}\sum_{\mathbf{k}}e^{-i(\mathbf{l}-\mathbf{m})\cdot\mathbf{k}}\left[x_{\mathbf{k}\sigma}^* x_{\mathbf{k}\sigma'}n_F(\epsilon_\mathbf{k}^-) + y_{\mathbf{k}\sigma}^* y_{\mathbf{k}\sigma'} n_F(\epsilon_\mathbf{k}^+)\right],
\end{equation}
where $\Omega$ is the volume of the full system and summation runs over $n$-dimensional Brillouin zone.

The main advantage with this approach over Eq.~\eqref{eq:cmat_general} is the ability to obtain explicit analytical expressions for eigenstates and eigenvalues in $k$-space. While the correlation matrix in itself is gauge invariant, care must be taken to ensure the gauge used in the eigenvectors is consistent. We will assume the Hamiltonians (pre- and post-quench) in $k$-space are of the form
\begin{equation}
    H^{i/f}_k = d^{i/f}_0\sigma_0 + \vec d^{i/f} \cdot \bs \sigma.
\end{equation}
where $i/f$ labels pre-quench (initial) and post-quench (final) parameters.
If it can be guaranteed that $d^{i/f}_3 \neq -d^{i/f} \equiv - |\vec d^{i/f}|$ for $k \in ]0,2\pi[$, a suitable gauge choice is
\begin{equation}
    \ket{u_{k+}} = \frac{1}{\sqrt{2d^{f}(d^{f}+d^{f}_3)}}
    \begin{pmatrix}
    d^{f}_3 + d^{f}\\ d^{f}_1 + id^{f}_2
    \end{pmatrix}
    \qquad
    \ket{u_{k-}} = \frac{1}{\sqrt{2d^{f}(d^{f}+d^{f}_3)}}
    \begin{pmatrix}d^{f}_1 - id^{f}_2\\-(d^{f}_3+d^{f})
    \end{pmatrix},
\end{equation}
where, as previously, $\ket{u_{k\pm}}$ is a post-quench eigenstate; the same holds for the pre-quench eigenstates $\ket{v_{k\pm}}$, but with $d^i$ instead.
This gauge works e.g. for the 1d topological insulator $\bs{d}(k)=(0,\sin k,0,m-\cos k)$ when $m > -1$. If, on the other hand, it is known that $m < 1$, we can instead use the gauge
\begin{equation}
    \ket{u_{k+}} = \frac{1}{\sqrt{2d^f(d^f-d^f_3)}}
    \begin{pmatrix}
    d^f_1 - id^f_2\\ d^f - d^f_3
    \end{pmatrix}
    \qquad
    \ket{u_{k-}} = \frac{1}{\sqrt{2d^f(d^f-d^f_3)}}
    \begin{pmatrix}d^f_3 - d^f\\d^f_1 + id^f_2
    \end{pmatrix}.
\end{equation}
For the Hamiltonians used in this work, these expressions are ill defined at $k = 0$, but an unambiguous limit exists and is be used instead.

\subsection{Partial translation invariance}

If the system is translationally invariant either post- or pre-quench but not both, a mixed $k$-space approach can be used. Eq.~\eqref{eq:cinkspace_final} can be applied directly, but by replacing the part that corresponds to the system which is not translationally invariant with the Fourier transform of its realspace solution. In other words, if the pre-quench system is not translation invariant, we substitute
\begin{equation}
	M_{k,k'} = \sum_\epsilon n_F(\epsilon)\ket{\phi^\epsilon_{k}}\bra{\phi^\epsilon_{k'}},
\end{equation}
while if the post-quench system breaks the invariance, we substitute
\begin{equation}
	\Psi_{n,\sigma,k}(t) = \sum_E \left[e^{-iEt}\psi_{n\sigma}^{E}\right]^*\ket{\psi^{E}_{k}}.
\end{equation}

In this way, the analytical expressions derived in the previous subsection can be used for the Hamiltonian that does preserve translation invariance. Explicitly, using the notation of the previous two cases:
\begin{enumerate}
\item  If the pre-quench system breaks translational symmetry, but the post-quench system does not, we have
\begin{equation}
	\mathcal C_{lm}^{\sigma\sigma'}(t) = \frac{1}{L}\sum_{k,k',\epsilon}e^{-ikl}x_{k,\sigma,\epsilon}^*(t)n_F(\epsilon)x_{k',\sigma',\epsilon}(t)e^{ik'm},
\end{equation}
where
\begin{equation}
	x_{k,\sigma,\epsilon}(t) = e^{-i\omega^+_kt} u_{k,\sigma,+}\langle\bs u_{k,+}|\phi_{\epsilon,k}\rangle + e^{-i\omega^-_kt} u_{k,\sigma,-}\langle\bs u_{k,-} |\phi_{\epsilon,k}\rangle.
\end{equation}

\item If the post-quench system breaks translational symmetry, while the pre-quench system does not, we have
 \begin{equation}
 	\mathcal C_{lm}^{\sigma\sigma'}(t) = \sum_{k}(\Psi^k_{m,\sigma'}(t))^\dagger   M_k \Psi^k_{l,\sigma}(t)
 \end{equation}

with 
$\Psi_{n,\sigma,k}(t) = \sum_E \left[e^{-iEt}\psi_{n\sigma}^{E}\right]^*\ket{\psi^{E}_{k}}$ and $ M_{k} = n_F(\epsilon_k^-) \ket{v_{k-}}  \bra{v_{k-}} + n_F(\epsilon_k^+) \ket{v_{k+}}  \bra{v_{k+}}$
\end{enumerate}

\subsection{Momentum-resolved dynamical correlation matrix}

The entanglement spectrum and the entanglement echo for translation invariant systems beyond 1d are conveniently analyzed as a function of momenta which are conserved by the entanglement cut defining the subsystem. For example, on a torus we can perform the subsystem partitioning in $x$ direction and maintain translation symmetry in perpendicular $y$ direction. The corresponding entanglement spectrum can be calculated from a correlation matrix when $k_y$ is a good quantum number. This $k_y$-resolved matrix is defined as 
$\mathcal{C}_{lm}^{\sigma\sigma'}(k_y,t)=\langle\hat{c}^\dagger_{l,k_y\sigma}(t)\hat{c}_{m,k_y\sigma'}(t)\rangle$ where $l,m$ label the reduced system coordinates in $x$ direction. Here the partially transformed fermion operators $\hat{c}_{m,k_y\sigma}$ annihilate particles at position $m$ with transverse momentum $k_y$ and spin $\sigma$. Following the derivation of the translation invariant two-band systems above, we obtain 
\begin{align}\label{eq:correlation 2d}
&\mathcal{C}_{lm}^{\sigma\sigma'}(k_y,t)=
\frac{1}{L_x}\sum_{k_x}e^{-i(l-m)k_x}\left[x_{k_x\sigma}^*(k_y,t) x_{k_x\sigma'}(k_y,t)n_F(\epsilon_{k_x}^-(k_y)) + y_{k_x\sigma}^*(k_y,t) y_{k_x\sigma'}(k_y,t) n_F(\epsilon_{k_x}^+(k_y))\right]
\end{align}
Here $L_x$ is the length of the full system in $x$ direction and the summation runs over all quasimomenta $k_x$.

Diagonalizing the matrix \eqref{eq:correlation 2d} for a fixed  $k_y$ provides eigenfunctions of the momentum-resolved single-particle entanglement Hamiltonian. The momentum-resolved entanglement echo is then obtained by the determinant formula Eq.~(3) in the main text. Repeating the calculation for each $k_y$ provides the full single-particle entanglement spectrum and all the partial entanglement echos.  


\section{Loschmidt echo}

For the completeness, here we derive the formula for the Loschmidt echo for general position-dependent quench. The Loschmidt echo is defined as $M(t) = |\mathcal L(t)|^2$, where $\mathcal L(t)$ is the Loschmidt amplitude

\begin{align} \label{eq:L1}
	\mathcal{L}(t)=\langle\Psi_0|U(t)|\Psi_0\rangle=\langle 0|\hat{a}_{\epsilon_1}\ldots\hat{a}_{\epsilon_{N-1}}\hat{a}_{\epsilon_N} U(t)\tilde{\hat{a}}_{\epsilon_N}^\dagger\hat{a}_{\epsilon_{N-1}}^\dagger\ldots\hat{a}_{\epsilon_1}^\dagger |0\rangle, 
\end{align}
where $\tilde{\hat{c}}_{\epsilon_i}^\dagger$ creates a particle in an eigenstate of the pre-quench Hamiltonian. We convert the operators to the post-quench basis using 
$\hat{a}_{\epsilon_{i}}(t)=\sum_{E_i}\langle \epsilon_i|E_i \rangle  \hat{c}_{E_i}$,
and get
\begin{align} 
	\mathcal{L}(t)= \sum_{E_i,E_i'}&\langle \epsilon_N|E_N'\rangle\langle E_N |\epsilon_N\rangle e^{-iE_N t}\times\ldots \times\langle \epsilon_1|E_1'\rangle\langle E_1 |\epsilon_1\rangle e^{-iE_1 t} \nonumber\\ 
	&\times\langle 0|\hat{c}_{E'_1}\ldots\hat{c}_{E'_{N-1}}U(t)\hat{c}_{E'_N}\hat{c}_{E_N}^\dagger\hat{c}_{E_{N-1}}^\dagger\ldots\hat{c}_{E_1}^\dagger |0\rangle.
\end{align}
By inserting $1 = U^\dagger(t)U(t)$ between each set of creation operators, we obtain a product of terms of the form $U(t)c^\dagger_{E_i} U^\dagger(t) = c^\dagger_{E_i}(-t)$, where the time dependence can then be easily obtained to get
\begin{align} \label{eq:L3}
	\mathcal{L}(t)=&\sum_{E_i,E_i'}\langle \epsilon_N|E_N'\rangle\langle E_N |\epsilon_N\rangle e^{-iE_N t}\times\ldots \times\langle \epsilon_1|E_1'\rangle\langle E_1 |\epsilon_1\rangle e^{-iE_1 t} \nonumber\\ 
	&\times\langle 0|\hat{c}_{E'_1}\ldots\hat{c}_{E'_{N-1}}\hat{c}_{E'_N}\hat{c}_{E_N}^\dagger\hat{c}_{E_{N-1}}^\dagger\ldots\hat{c}_{E_1}^\dagger |0\rangle.
\end{align}
The operator expectation value vanishes unless the set $E_n'$ is a permutation of the set $E_n$ in which case it gives the sign of the permutation. Thus, we can write the amplitude as a sum over the set $E_i$ and permutations $P_i$ as  
\begin{align} \label{eq:L3}
	\mathcal{L}(t)=&\sum_{E_i,P_i} \mathrm{sign}P_i \langle \epsilon_N|E_{P_iN}\rangle\langle E_N |\epsilon_N\rangle e^{-iE_N t}\times\ldots \times\langle \epsilon_1|E_{P_i1}\rangle\langle E_1 |\epsilon_1\rangle e^{-iE_1 t},
\end{align}
which remains unchanged if we switch the permutation in the ket vectors into the bra vectors  
\begin{align} \label{eq:L4}
	\mathcal{L}(t)=&\sum_{E_i,P_i} \mathrm{sign}P_i \langle \epsilon_{P_iN}|E_{N}\rangle\langle E_N |\epsilon_N\rangle e^{-iE_N t}\times\ldots \times\langle \epsilon_{P_i1}|E_{1}\rangle\langle E_1 |\epsilon_1\rangle e^{-iE_1 t}.
\end{align}
Finally, this can be written as 
\begin{align} \label{eq:Loschmidt_inhomogeneous}
	\mathcal{L}(t)=\det{\mathcal{A}},   
\end{align}
where 
\begin{align} \label{eq:A}
	\left[\mathcal{A}\right]_{ij}=\sum_{E} \langle \epsilon_{i}|E\rangle  \langle E|\epsilon_j\rangle e^{-iE t}.   
\end{align}
In the above, the label $E$ refers to the post-quench Hamiltonian while $\epsilon$ refers to the pre-quench $H$.
Typically rather than the echo $M(t) = |\mathcal{L}(t)|^2$, one studies the Loschmidt rate function given by 
\begin{equation}
	\lambda(t)=-\log{|\mathcal{L}(t)|^2}/\Omega,
\end{equation}
where $\Omega$ is the volume of the system. Hence, zeros of the Loschmidt echo correspond to divergences in the rate function.

The general formula above simplifies for two-band systems. When the pre-quench Hamiltonian has a filled lower-band $|\phi_{\epsilon_i}^-\rangle$ and the post-quench Hamiltonian is translation invariant, we have 
\begin{align} 
	\left[\mathcal{A}\right]_{ij}=\sum_{k}\left[\langle{\phi_{\epsilon_i}^-}|u_k^+\rangle\langle{ u_k^+}|\phi_{\epsilon_j}^-\rangle e^{i\omega_k^-t}+\langle{\phi_{\epsilon_i}^-}|u_k^-\rangle\langle{u_k^-}|\phi_{\epsilon_j}^-\rangle e^{i\omega_k^+t}  \right],    
\end{align}
where the summation is over the momentum eigenstates, and, again, $|u_k^\pm\rangle$ are eigenstates of the post-quench Hamiltonian. Furthermore, if also the initial state is translation invariant,  the general formula reduces to the well-known form
\begin{align} \label{Loschmidt}
	\mathcal{L}(t)=\prod_{k} \left[\cos(d^f_kt)+i\sin(d^f_kt)\cos\theta_k \right],   
\end{align}
where $\theta_k$ is the polar angle between $\vec d^i$ and $\vec d^f$ on the Bloch sphere and the product is over $n$-dimensional Brillouin zone.


\section{Critical times of entanglement transitions}

\begin{figure}
	\centering
	\includegraphics[width=0.99\columnwidth]{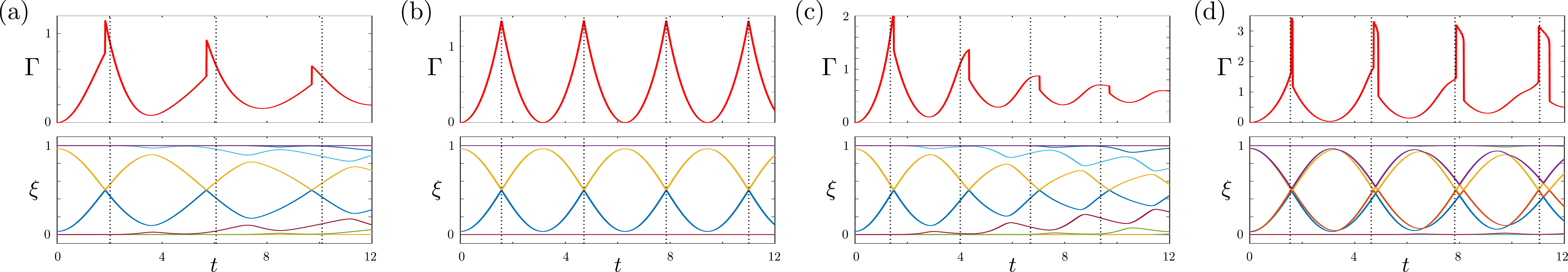}
	\caption{(a): Entanglement echo rate (top) and spectrum (bottom) of 1d system for the quench $m=1.5 \to m=0.3$ and system size $L=100$ (total) $L_A=30$ (subsystem). The dashed vertical lines indicate Loschmidt critical times. (b) Same, but for the quench $m=1.5 \to m=0.0$. (c) Same, but for the quench $m=1.5 \to m=-0.3$. (d) Same, but where $m$ on each site is picked from a uniform distribution $m = 1.5 \pm 0.3$ pre-quench and $m = \pm 0.3$ post-quench.}
	\label{suppfigtimes}
\end{figure}

As noted in the main text, while the critical times of bulk-type transitions coincide with those obtained from the Loschmidt amplitude, this is not the case for entanglement-type transition. In fact, this deviation from the Loschmidt critical times is a parameter-dependent phenomenon, as seen from Fig.~\ref{suppfigtimes}. For $m > 0$ the entanglement transition occurs with shorter critical time, with the opposite for negative $m$; at the special point $m = 0$, the entanglement critical time coincides with the Loschmidt value.
While we have kept the pre-quench $m$ constant in the figure, this occurs even if it is varied e.g. so that each quench corresponds to a constant shift in $m$ instead. Generically, then, an entanglement-type topological transition will have a different critical time than would the analogous Loschmidt transition. However, the precise shift depend on the details of the quench; in fact, as seen in Fig.~\ref{suppfigtimes}(d), when disorder is included, the critical times of entanglement transitions split due to the lifting of the translation symmetry protected four-fold degeneracy of the midgap crossings to two different crossings. Regardless, the discontinuities in the entanglement echo and the crossings in the entanglement spectrum remain as telltale signs of an entanglement-type transitions.

\section{particle number fluctuations}
Here we will derive the formula for the time-dependent fluctuations of the number of particles in the subsystem $A$ in terms of the single-particle entanglement spectrum evaluated in a gaussian pre-quench state. The starting point is the particle number operator of the subsystem $A$
\begin{align} \label{eq:particle1}
\hat{N}_A=\sum_{i\in A,\sigma=\uparrow,\downarrow}  \hat{c}_{i\sigma}^\dagger\hat{c}_{i\sigma},
\end{align}
with expectation value $\langle\hat{N}_A\rangle(t)=\sum_{i\in A,\sigma=\uparrow,\downarrow} \TR_A[\rho_A(t)\hat{c}_{i\sigma}^\dagger\hat{c}_{i\sigma}]=\sum_{i\in A,\sigma=\uparrow,\downarrow} \langle\hat{c}_{i\sigma}^\dagger\hat{c}_{i\sigma}\rangle=\mathrm{tr}\,{\mathcal{C}}(t)=\sum_{i}\xi_i(t)$. This quantity is fixed by the total number of particles in a homogeneous system. However, the subsystem particle number fluctuations display a more interesting behaviour as they are sensitive to  non-local correlations over the whole subsystem. The variance of the particle number is obtained through 
\begin{align} \label{eq:variance}
&\mathrm{Var}\, N_A(t)=\langle\hat{N}_A^2\rangle-\langle\hat{N}_A\rangle^2= \sum_{\substack{i,j\in A \\ \sigma,\sigma'=\uparrow,\downarrow}}  \langle\hat{c}_{i\sigma}^\dagger\hat{c}_{i\sigma}\hat{c}_{j\sigma'}^\dagger\hat{c}_{j\sigma'}\rangle-\langle\hat{N}_A\rangle^2\nonumber\\
&=\sum_{\substack{i,j\in A \\ \sigma,\sigma'=\uparrow,\downarrow}} \langle\hat{c}_{i\sigma}^\dagger\hat{c}_{i\sigma}\rangle\langle\hat{c}_{j\sigma'}^\dagger\hat{c}_{j\sigma'}\rangle+ \langle\hat{c}_{i\sigma}^\dagger\hat{c}_{j\sigma'}\rangle\langle\hat{c}_{i\sigma}\hat{c}_{j\sigma'}^\dagger\rangle 
-\langle\hat{N}_A\rangle^2= \sum_{\substack{i,j\in A \\ \alpha,\sigma'=\uparrow,\downarrow}} \langle\hat{c}_{i\sigma}^\dagger\hat{c}_{j\sigma'}\rangle\langle\hat{c}_{i\sigma}\hat{c}_{j\sigma'}^\dagger\rangle, 
\end{align}
where Wick's theorem was employed to factor the four-operator expectation value. Thus, 
\begin{align} \label{eq:variance2}
\mathrm{Var}\,N_A(t)&=\sum_{\substack{i,j\in A \\ \sigma,\sigma'=\uparrow,\downarrow}} \langle\hat{c}_{i\sigma}^\dagger\hat{c}_{j\sigma'}\rangle\langle\hat{c}_{i\sigma}\hat{c}_{j\sigma'}^\dagger\rangle=\sum_{\substack{i,j\in A \\ \sigma,\sigma'=\uparrow,\downarrow}} \langle\hat{c}_{i\sigma}^\dagger\hat{c}_{j\sigma'}\rangle\left(\delta_{ij}^{\sigma\sigma'}-\langle\hat{c}_{j\sigma'}^\dagger\hat{c}_{i\sigma}\rangle\right)\notag\\
&=\mathrm{tr}\,{\mathcal{C}}(t)-\mathrm{tr}\,\mathcal{C}^2(t)=\sum_{i}\left[\xi_i(t)-\xi_i^2(t)\right] 
\end{align}
The final expression \eqref{eq:variance2} is valid for zero- as well as finite-temperature initial states and provides a direct connection between the dynamical entanglement spectrum and experimentally accessible quantities.

\section{Entanglement-type dynamical phase transitions at finite temperature}

\begin{figure*}[ht]
    \centering
    \includegraphics[width=0.99\columnwidth]{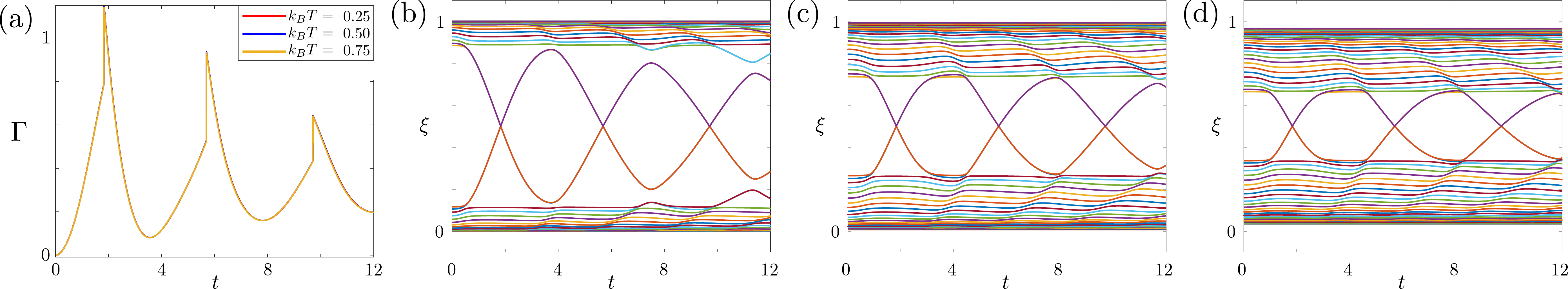}
    \caption{(a): Entanglement echo rate of 1d system (top) for quench $m=1.5 \to m=0.3$ and the system size $L=100$ (total) $L_A=30$ (subsystem). (b)-(d): Corresponding single particle entanglement spectra in the order of increasing temperature with $k_BT =$  0.25, 0.5 and 0.75 in units of $t$. }
    \label{suppfig1}
\end{figure*}

In this section we will illustrate the behaviour of the entanglement echo for finite-temperature initial states. In particular, it is shown that the entanglement echo still predicts a sharp transition which has a clear physical interpretation. In Fig.~\ref{suppfig1} (a) we have plotted the entanglement echo for a quench displaying entanglement-type dynamical transitions and in Figs.~\ref{suppfig1} (b)-(d) the corresponding single-particle entanglement spectra. Notably, the echo rate is essentially temperature independent, displaying jump singularities coinciding with the instantaneous entanglement ground state degeneracies arising from the level crossings at $\xi=1/2$. Increasing temperature primarily serves to compress the edges of the entanglement spectrum towards $\xi = 1/2$. Hence, the crossing are robust and do not shift when the temperature of the initial state is varied, preserving the discontinuities in the echo rate. 

\begin{figure}[ht]
    \centering
    \includegraphics[width=0.99\columnwidth]{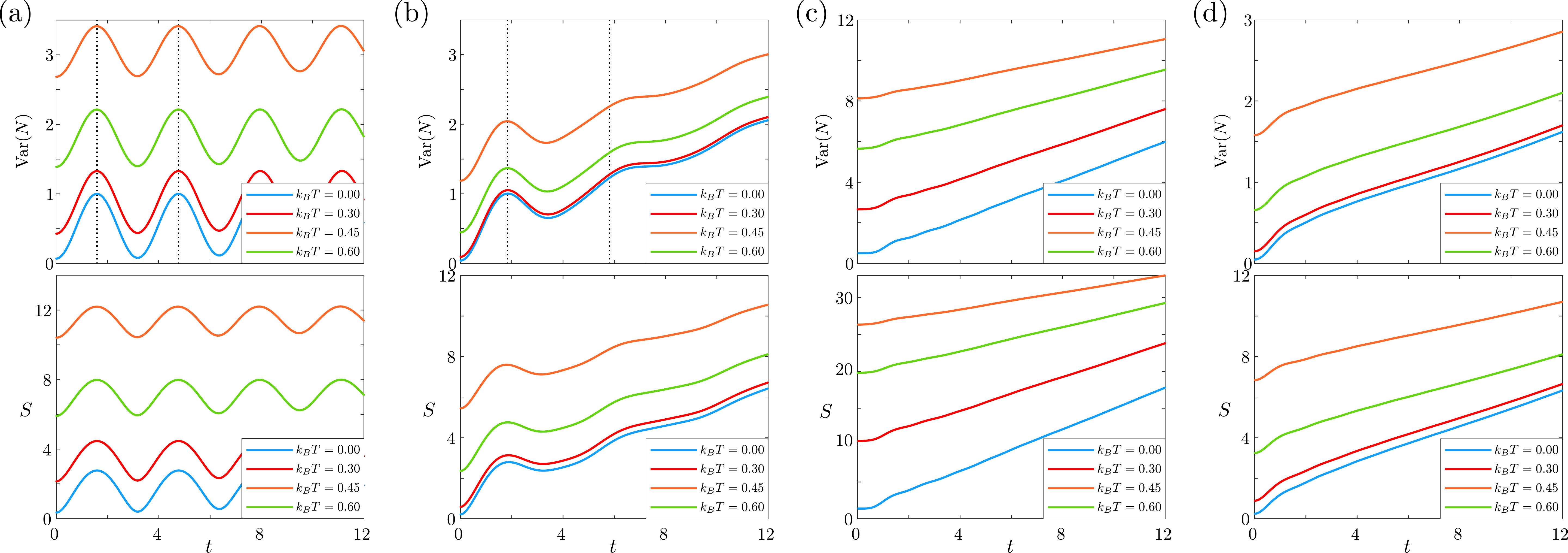}
    \caption{Particle number fluctuations (top) and the entanglement entropy (bottom). (a): Entanglement-type transition $m=2.0 \to m=0.05$. The dashed line indicates first two critical times. (b): Entanglement transition $m=2.6 \to m=0.4$  (c): bulk-type transition $m=-0.3 \to m=1.6$ (d): trivial quench $m=2.4 \to m=1.2$. }
    \label{suppfig2}
\end{figure}

The fact that the non-analytic character of entanglement-type transitions  persists to finite temperatures has observable physical consequences for the subsystem fluctuations, as discussed in the main text. The time-dependent variance of the particle number of a subsystem, illustrated in Fig.~\ref{suppfig2}, shows how the oscillating entanglement ground state degeneracy translates to pronounced oscillations in entanglement-type quenches shown in (a) and (b). The frequency of the oscillations reflects the periodicity of the critical times, while the onset of the linear trend depends on the particular post-quench parameters. In the special case where the post-quench Hamiltonian is tuned close to $m=0$, the oscillations persist a remarkably long time before the linear increase sets in. The entanglement spectrum crossings correspond to local maxima in the particle number variance, which provides an experimental handle to extract critical times. Due to the four-fold degeneracy of crossings in the model, the maximum variance at $T=0$ is quantized to 1, as can be observed at the critical times before the onset of the approximately linear growth. An increased initial temperature serves to shift the overall variance upwards, which obscures the quantization; however, the overall oscillatory behaviour, while somewhat lower in amplitude, can still clearly be observed. Remarkably, this behaviour of the particle number fluctuations is essentially the same as that of the von Neumann entropy, which is not easily accessed experimentally. 

In comparison to the entanglement type transitions, the particle number and entropy oscillations are suppressed for bulk-type transitions and trivial quenches, as illustrated in Fig.~\ref{suppfig2} (c) and (d). This difference could be employed as an experimental probe to distinguish entanglement-type transitions from bulk transition and trivial quenches.

\end{document}